\documentclass[acmsmall]{acmart}

\acmJournal{PACMPL}
\acmVolume{1}
\acmNumber{CONF} 
\acmArticle{1}
\acmYear{2018}
\acmMonth{1}
\acmDOI{} 
\startPage{1}

\setcopyright{none}

\usepackage{amsmath}
\usepackage[shortlabels]{enumitem}
\usepackage{amssymb}
\usepackage{ifthen}
\usepackage{amsfonts}
\usepackage{float}
\usepackage{ifthen}
\usepackage{multicol}
\usepackage{multirow}
\usepackage{xcolor}
\usepackage{framed}
\usepackage{caption}
\usepackage{tcolorbox}
\usepackage{tikz}
\usepackage{xspace}
\usepackage{amsthm}
\usepackage{thmtools}
\usepackage{mathpartir}
\usepackage{xparse}
\usepackage{stmaryrd}
\usepackage{environ}
\usepackage{natbib}
\usepackage{cancel}
\usepackage{hyperref}
\usepackage{cleveref}
\usepackage{hhline}

\usetikzlibrary{matrix}
\usetikzlibrary{decorations.pathreplacing,angles,quotes}
\usetikzlibrary{calc}
\usetikzlibrary{backgrounds}
\usetikzlibrary{positioning}

\tikzset{
  nodraw/.style={draw=none,fill=none}
}
\usetikzlibrary{decorations}
\pgfdeclaredecoration{arrows}{draw}{
\state{draw}[width=\pgfdecoratedinputsegmentlength]{%
  \path [every arrow subpath/.try] \pgfextra{%
    \pgfpathmoveto{\pgfpointdecoratedinputsegmentfirst}%
    \pgfpathlineto{\pgfpointdecoratedinputsegmentlast}%
   };
}}
\tikzset{every arrow subpath/.style={->, draw, thick}}

\definecolor{light-gray}{gray}{0.85}
\definecolor{shadecolor}{gray}{0.85}

\usepackage{booktabs}   
\usepackage{subcaption} 

\input{macros}

\newcommand{\fsub}{F_{<:}\xspace}
\newcommand{\fsubsup}{F_{<:>}\xspace}
\newcommand{\fsubm}{\fsub^-}

\newcommand{\dsub}{D_{<:}\xspace}
\newcommand{\dint}{D_{\wedge}\xspace}
\newcommand{\dsubsub}{D_{<<:}\xspace}
\newcommand{\DOT}{DOT\xspace}
\newcommand{\jDOT}{jDOT\xspace}
\newcommand{\mdart}{\mu{DART}\xspace}

\newcommand\tthier{\text{tdh}}

\DeclareDocumentCommand{\typing}{ o m m } {
  \IfNoValueTF {#1}
  {\Gamma \vdash #2 : #3}
  {#1 \vdash #2 : #3}
}

\DeclareDocumentCommand{\subtyping}{ o m m } {
  \IfNoValueTF {#1}
  {\Gamma \vdash #2 <: #3}
  {#1 \vdash #2 <: #3}
}

\DeclareDocumentCommand{\subss}{ o m m } {
  \IfNoValueTF {#1}
  {\Gamma \vdash #2 \uparrow #3}
  {#1 \vdash #2 \uparrow #3}
}

\DeclareDocumentCommand{\typingdot}{ o m m } {
  \IfNoValueTF {#1}
  {\Gamma \vdash_{\DOT} #2 : #3}
  {#1 \vdash_{\DOT} #2 : #3}
}

\DeclareDocumentCommand{\subtypingdot}{ o m m } {
  \IfNoValueTF {#1}
  {\Gamma \vdash_{\DOT} #2 <: #3}
  {#1 \vdash_{\DOT} #2 <: #3}
}

\DeclareDocumentCommand{\typingd}{ o m m } {
  \IfNoValueTF {#1}
  {\Gamma \vdash_{\dsub} #2 : #3}
  {#1 \vdash_{\dsub} #2 : #3}
}

\DeclareDocumentCommand{\subtypingd}{ o m m } {
  \IfNoValueTF {#1}
  {\Gamma \vdash_{\dsub} #2 <: #3}
  {#1 \vdash_{\dsub} #2 <: #3}
}

\DeclareDocumentCommand{\subtypingdp}{ o m m } {
  \IfNoValueTF {#1}
  {\Gamma \vdash_{\dsub^+} #2 <: #3}
  {#1 \vdash_{\dsub^+} #2 <: #3}
}

\DeclareDocumentCommand{\subtypingdss}{ o m m } {
  \IfNoValueTF {#1}
  {\Gamma \vdash_{\dsubsub} #2 <: #3}
  {#1 \vdash_{\dsubsub} #2 <: #3}
}

\DeclareDocumentCommand{\subtypingf}{ o m m } {
  \IfNoValueTF {#1}
  {\Gamma \vdash_{\fsub} #2 <: #3}
  {#1 \vdash_{\fsub} #2 <: #3}
}

\DeclareDocumentCommand{\subtypingfd}{ o m m } {
  \IfNoValueTF {#1}
  {\Gamma \vdash_{\fsub^d} #2 <: #3}
  {#1 \vdash_{\fsub^d} #2 <: #3}
}

\DeclareDocumentCommand{\subtypingfm}{ o m m } {
  \IfNoValueTF {#1}
  {\Gamma \vdash_{\fsub^-} #2 <: #3}
  {#1 \vdash_{\fsub^-} #2 <: #3}
}

\DeclareDocumentCommand{\subtypingfmn}{ o m m } {
  \IfNoValueTF {#1}
  {\Gamma \vdash_{\fsub^{-NF}} #2 <: #3}
  {#1 \vdash_{\fsub^{-NF}} #2 <: #3}
}

\DeclareDocumentCommand{\subssfm}{ o m m } {
  \IfNoValueTF {#1}
  {\Gamma \vdash_{\fsub^-} #2 \uparrow #3}
  {#1 \vdash_{\fsub^-} #2 \uparrow #3}
}

\DeclareDocumentCommand{\subssfms}{ o m m } {
  \IfNoValueTF {#1}
  {\Gamma \vdash_{\fsub^-} #2 \uparrow^* #3}
  {#1 \vdash_{\fsub^-} #2 \uparrow^* #3}
}

\DeclareDocumentCommand{\subtypingfsm}{ o m m } {
  \IfNoValueTF {#1}
  {\Gamma \vdash_{\fsubsup^-} #2 <: #3}
  {#1 \vdash_{\fsubsup^-} #2 <: #3}
}

\DeclareDocumentCommand{\subssfsm}{ o m m } {
  \IfNoValueTF {#1}
  {\Gamma \vdash_{\fsubsup^-} #2 \uparrow #3}
  {#1 \vdash_{\fsubsup^-} #2 \uparrow #3}
}

\DeclareDocumentCommand{\subssfsms}{ o m m } {
  \IfNoValueTF {#1}
  {\Gamma \vdash_{\fsubsup^-} #2 \uparrow^* #3}
  {#1 \vdash_{\fsubsup^-} #2 \uparrow^* #3}
}

\DeclareDocumentCommand{\subssd}{ o m m } {
  \IfNoValueTF {#1}
  {\Gamma \vdash_{\dsub} #2 \uparrow #3}
  {#1 \vdash_{\dsub} #2 \uparrow #3}
}

\DeclareDocumentCommand{\subssds}{ o m m } {
  \IfNoValueTF {#1}
  {\Gamma \vdash_{\dsub} #2 \uparrow^* #3}
  {#1 \vdash_{\dsub} #2 \uparrow^* #3}
}

\DeclareDocumentCommand{\expod}{ o m m } {
  \IfNoValueTF {#1}
  {\Gamma \vdash_{\dsub{S}} #2 \Uparrow #3}
  {#1 \vdash_{\dsub{S}} #2 \Uparrow #3}
}

\DeclareDocumentCommand{\uced}{ o m m } {
  \IfNoValueTF {#1}
  {\Gamma \vdash_{\dsub{S}} #2 \nearrow #3}
  {#1 \vdash_{\dsub{S}} #2 \nearrow #3}
}

\DeclareDocumentCommand{\dced}{ o m m } {
  \IfNoValueTF {#1}
  {\Gamma \vdash_{\dsub{S}} #2 \searrow #3}
  {#1 \vdash_{\dsub{S}} #2 \searrow #3}
}

\DeclareDocumentCommand{\udced}{ o m m m } {
  \IfNoValueTF {#1}
  {\Gamma \vdash_{\dsub{S}} #2 \nearrow #3 (\searrow #4)}
  {#1 \vdash_{\dsub{S}} #2 \nearrow #3 (\searrow #4)}
}

\DeclareDocumentCommand{\subtypingda}{ o m m } {
  \IfNoValueTF {#1}
  {\Gamma \vdash_{\dsub{S}} #2 <: #3}
  {#1 \vdash_{\dsub{S}} #2 <: #3}
}

\DeclareDocumentCommand{\typingda}{ o m m } {
  \IfNoValueTF {#1}
  {\Gamma \vdash_{\dsub{S}} #2 : #3}
  {#1 \vdash_{\dsub{S}} #2 : #3}
}

\DeclareDocumentCommand{\promod}{ o m m m } {
  \IfNoValueTF {#1}
  {\Gamma \vdash_{\dsub{S}} #2 \Uparrow_{#3} #4}
  {#1 \vdash_{\dsub{S}} #2 \Uparrow_{#3} #4}
}

\DeclareDocumentCommand{\demod}{ o m m m } {
  \IfNoValueTF {#1}
  {\Gamma \vdash_{\dsub{S}} #2 \Downarrow_{#3} #4}
  {#1 \vdash_{\dsub{S}} #2 \Downarrow_{#3} #4}
}

\DeclareDocumentCommand{\promdemod}{ o m m m } {
  \IfNoValueTF {#1}
  {\Gamma \vdash_{\dsub{S}} #2 \Uparrow_{#3}(\Downarrow_{#3}) #4}
  {#1 \vdash_{\dsub{S}} #2 \Uparrow_{#3}(\Downarrow_{#3}) #4}
}

\DeclareDocumentCommand{\dempromod}{ o m m m } {
  \IfNoValueTF {#1}
  {\Gamma \vdash_{\dsub{S}} #2 \Downarrow_{#3}(\Uparrow_{#3}) #4}
  {#1 \vdash_{\dsub{S}} #2 \Downarrow_{#3}(\Uparrow_{#3}) #4}
}

\DeclareDocumentCommand{\stareat}{ o m m o } {
  \IfNoValueTF {#1}
  {
    \IfNoValueTF {#4}
    { \Gamma_1 \gg  #2 <: #3 \ll \Gamma_2 }
    { \Gamma_1 \gg  #2 <: #3 \ll #4 }
  }
  {
    \IfNoValueTF {#4}
    { #1 \gg  #2 <: #3 \ll \Gamma_2 }
    { #1 \gg  #2 <: #3 \ll #4 }
  } 
}

\DeclareDocumentCommand{\stareatr}{ o m m o } {
  \IfNoValueTF {#1}
  {
    \IfNoValueTF {#4}
    { \Gamma_1 \gg  #2 >: #3 \ll \Gamma_2 }
    { \Gamma_1 \gg  #2 >: #3 \ll #4 }
  }
  {
    \IfNoValueTF {#4}
    { #1 \gg  #2 >: #3 \ll \Gamma_2 }
    { #1 \gg  #2 >: #3 \ll #4 }
  } 
}

\DeclareDocumentCommand{\ucd}{ o m m o } {
  \IfNoValueTF {#1}
  {
    \IfNoValueTF {#4}
    { \Gamma \vdash_{\dsub{S}} #2 \nearrow #3 \dashv \Gamma' }
    { \Gamma \vdash_{\dsub{S}} #2 \nearrow #3 \dashv #4 }
  }
  {
    \IfNoValueTF {#4}
    { #1 \vdash_{\dsub{S}} #2 \nearrow #3 \dashv #1' }
    { #1 \vdash_{\dsub{S}} #2 \nearrow #3 \dashv #4 }
  } 
}

\DeclareDocumentCommand{\dcd}{ o m m o } {
  \IfNoValueTF {#1}
  {
    \IfNoValueTF {#4}
    { \Gamma \vdash_{\dsub{S}} #2 \searrow #3 \dashv \Gamma' }
    { \Gamma \vdash_{\dsub{S}} #2 \searrow #3 \dashv #4 }
  }
  {
    \IfNoValueTF {#4}
    { #1 \vdash_{\dsub{S}} #2 \searrow #3 \dashv #1' }
    { #1 \vdash_{\dsub{S}} #2 \searrow #3 \dashv #4 }
  } 
}

\DeclareDocumentCommand{\udcd}{ o m m o } {
  \IfNoValueTF {#1}
  {
    \IfNoValueTF {#4}
    { \Gamma \vdash_{\dsub{S}} #2 \nearrow(\searrow) #3 \dashv \Gamma' }
    { \Gamma \vdash_{\dsub{S}} #2 \nearrow(\searrow) #3 \dashv #4 }
  }
  {
    \IfNoValueTF {#4}
    { #1 \vdash_{\dsub{S}} #2 \nearrow(\searrow) #3 \dashv #1' }
    { #1 \vdash_{\dsub{S}} #2 \nearrow(\searrow) #3 \dashv #4 }
  } 
}

\DeclareDocumentCommand{\revealingd}{ o m m o } {
  \IfNoValueTF {#1}
  {
    \IfNoValueTF {#4}
    { \Gamma \vdash_{\dsub{S}} #2 \Uuparrow #3 \dashv \Gamma' }
    { \Gamma \vdash_{\dsub{S}} #2 \Uuparrow #3 \dashv #4 }
  }
  {
    \IfNoValueTF {#4}
    { #1 \vdash_{\dsub{S}} #2 \Uuparrow #3 \dashv #1' }
    { #1 \vdash_{\dsub{S}} #2 \Uuparrow #3 \dashv #4 }
  } 
}

\DeclareDocumentCommand{\subtypingdk}{ o m m } {
  \IfNoValueTF {#1}
  {\Gamma \vdash_{\dsub{K}} #2 <: #3}
  {#1 \vdash_{\dsub{K}} #2 <: #3}
}

\DeclareDocumentCommand{\subtypingdt}{ o m m o } {
  \IfNoValueTF {#1}
  {
    \IfNoValueTF {#4}
    { \Gamma \vdash_{\dsub} #2 <: #3 \dashv \Gamma }
    { \Gamma \vdash_{\dsub} #2 <: #3 \dashv #4 }
  }
  {
    \IfNoValueTF {#4}
    { #1 \vdash_{\dsub} #2 <: #3 \dashv \Gamma }
    { #1 \vdash_{\dsub} #2 <: #3 \dashv #4 }
  } 
}

\DeclareDocumentCommand{\subtypingdsk}{ o m m o } {
  \IfNoValueTF {#1}
  {
    \IfNoValueTF {#4}
    { \Gamma_1 \vdash_{\dsub{SK}} #2 <: #3 \dashv \Gamma_2 }
    { \Gamma_1 \vdash_{\dsub{SK}} #2 <: #3 \dashv #4 }
  }
  {
    \IfNoValueTF {#4}
    { #1 \vdash_{\dsub{SK}} #2 <: #3 \dashv \Gamma_2 }
    { #1 \vdash_{\dsub{SK}} #2 <: #3 \dashv #4 }
  } 
}

\DeclareDocumentCommand{\subtypingdskr}{ o m m o } {
  \IfNoValueTF {#1}
  {
    \IfNoValueTF {#4}
    { \Gamma_1 \vdash_{\dsub{SK}} #2 >: #3 \dashv \Gamma_2 }
    { \Gamma_1 \vdash_{\dsub{SK}} #2 >: #3 \dashv #4 }
  }
  {
    \IfNoValueTF {#4}
    { #1 \vdash_{\dsub{SK}} #2 >: #3 \dashv \Gamma_2 }
    { #1 \vdash_{\dsub{SK}} #2 >: #3 \dashv #4 }
  } 
}

\newcommand\exposure{\textbf{Exposure}\xspace}

\newcommand\upcast{\textbf{Upcast}\xspace}
\newcommand\downcast{\textbf{Downcast}\xspace}
\newcommand\revealing{\textbf{Revealing}\xspace}
\newcommand\opesub{OPE_{<:}\xspace}

\DeclareDocumentCommand{\opesubd}{ m o } {
  \IfNoValueTF {#2}
  {#1 \subseteq_{<:} #1'}
  {#1 \subseteq_{<:} #2}
}

\DeclareDocumentCommand{\subtypingdi}{ o m m } {
  \IfNoValueTF {#1}
  {\Gamma \vdash_{\dint} #2 <: #3}
  {#1 \vdash_{\dint} #2 <: #3}
}

\DeclareDocumentCommand{\ucdi}{ o m m o } {
  \IfNoValueTF {#1}
  {
    \IfNoValueTF {#4}
    { \Gamma \vdash_{\dint{S}} #2 \nearrow #3 \dashv \Gamma' }
    { \Gamma \vdash_{\dint{S}} #2 \nearrow #3 \dashv #4 }
  }
  {
    \IfNoValueTF {#4}
    { #1 \vdash_{\dint{S}} #2 \nearrow #3 \dashv #1' }
    { #1 \vdash_{\dint{S}} #2 \nearrow #3 \dashv #4 }
  } 
}

\DeclareDocumentCommand{\dcdi}{ o m m o } {
  \IfNoValueTF {#1}
  {
    \IfNoValueTF {#4}
    { \Gamma \vdash_{\dint{S}} #2 \searrow #3 \dashv \Gamma' }
    { \Gamma \vdash_{\dint{S}} #2 \searrow #3 \dashv #4 }
  }
  {
    \IfNoValueTF {#4}
    { #1 \vdash_{\dint{S}} #2 \searrow #3 \dashv #1' }
    { #1 \vdash_{\dint{S}} #2 \searrow #3 \dashv #4 }
  } 
}

\DeclareDocumentCommand{\udcdi}{ o m m o } {
  \IfNoValueTF {#1}
  {
    \IfNoValueTF {#4}
    { \Gamma \vdash_{\dint{S}} #2 \nearrow(\searrow) #3 \dashv \Gamma' }
    { \Gamma \vdash_{\dint{S}} #2 \nearrow(\searrow) #3 \dashv #4 }
  }
  {
    \IfNoValueTF {#4}
    { #1 \vdash_{\dint{S}} #2 \nearrow(\searrow) #3 \dashv #1' }
    { #1 \vdash_{\dint{S}} #2 \nearrow(\searrow) #3 \dashv #4 }
  } 
}

\DeclareDocumentCommand{\revealingdi}{ o m m o } {
  \IfNoValueTF {#1}
  {
    \IfNoValueTF {#4}
    { \Gamma \vdash_{\dint{S}} #2 \Uuparrow #3 \dashv \Gamma' }
    { \Gamma \vdash_{\dint{S}} #2 \Uuparrow #3 \dashv #4 }
  }
  {
    \IfNoValueTF {#4}
    { #1 \vdash_{\dint{S}} #2 \Uuparrow #3 \dashv #1' }
    { #1 \vdash_{\dint{S}} #2 \Uuparrow #3 \dashv #4 }
  } 
}

\DeclareDocumentCommand{\subtypingdisk}{ o m m o } {
  \IfNoValueTF {#1}
  {
    \IfNoValueTF {#4}
    { \Gamma_1 \vdash_{\dint{SK}} #2 <: #3 \dashv \Gamma_2 }
    { \Gamma_1 \vdash_{\dint{SK}} #2 <: #3 \dashv #4 }
  }
  {
    \IfNoValueTF {#4}
    { #1 \vdash_{\dint{SK}} #2 <: #3 \dashv \Gamma_2 }
    { #1 \vdash_{\dint{SK}} #2 <: #3 \dashv #4 }
  } 
}

\DeclareDocumentCommand{\subtypingdiskr}{ o m m o } {
  \IfNoValueTF {#1}
  {
    \IfNoValueTF {#4}
    { \Gamma_1 \vdash_{\dint{SK}} #2 >: #3 \dashv \Gamma_2 }
    { \Gamma_1 \vdash_{\dint{SK}} #2 >: #3 \dashv #4 }
  }
  {
    \IfNoValueTF {#4}
    { #1 \vdash_{\dint{SK}} #2 >: #3 \dashv \Gamma_2 }
    { #1 \vdash_{\dint{SK}} #2 >: #3 \dashv #4 }
  } 
}

\DeclareDocumentCommand{\typingmd}{ o m m } {
  \IfNoValueTF {#1}
  {\Gamma \vdash_{\mdart} #2 : #3}
  {#1 \vdash_{\mdart} #2 : #3}
}

\DeclareDocumentCommand{\subtypingmd}{ o m m } {
  \IfNoValueTF {#1}
  {\Gamma \vdash_{\mdart} #2 <: #3}
  {#1 \vdash_{\mdart} #2 <: #3}
}

\DeclareDocumentCommand{\ucmd}{ o m m o } {
  \IfNoValueTF {#1}
  {
    \IfNoValueTF {#4}
    { \Gamma \vdash_{\mdart{S}} #2 \nearrow #3 \dashv \Gamma' }
    { \Gamma \vdash_{\mdart{S}} #2 \nearrow #3 \dashv #4 }
  }
  {
    \IfNoValueTF {#4}
    { #1 \vdash_{\mdart{S}} #2 \nearrow #3 \dashv #1' }
    { #1 \vdash_{\mdart{S}} #2 \nearrow #3 \dashv #4 }
  } 
}

\DeclareDocumentCommand{\dcmd}{ o m m o } {
  \IfNoValueTF {#1}
  {
    \IfNoValueTF {#4}
    { \Gamma \vdash_{\mdart{S}} #2 \searrow #3 \dashv \Gamma' }
    { \Gamma \vdash_{\mdart{S}} #2 \searrow #3 \dashv #4 }
  }
  {
    \IfNoValueTF {#4}
    { #1 \vdash_{\mdart{S}} #2 \searrow #3 \dashv #1' }
    { #1 \vdash_{\mdart{S}} #2 \searrow #3 \dashv #4 }
  } 
}

\DeclareDocumentCommand{\udcmd}{ o m m o } {
  \IfNoValueTF {#1}
  {
    \IfNoValueTF {#4}
    { \Gamma \vdash_{\mdart{S}} #2 \nearrow(\searrow) #3 \dashv \Gamma' }
    { \Gamma \vdash_{\mdart{S}} #2 \nearrow(\searrow) #3 \dashv #4 }
  }
  {
    \IfNoValueTF {#4}
    { #1 \vdash_{\mdart{S}} #2 \nearrow(\searrow) #3 \dashv #1' }
    { #1 \vdash_{\mdart{S}} #2 \nearrow(\searrow) #3 \dashv #4 }
  } 
}

\DeclareDocumentCommand{\revealingmd}{ o m m o } {
  \IfNoValueTF {#1}
  {
    \IfNoValueTF {#4}
    { \Gamma \vdash_{\mdart{S}} #2 \Uuparrow #3 \dashv \Gamma' }
    { \Gamma \vdash_{\mdart{S}} #2 \Uuparrow #3 \dashv #4 }
  }
  {
    \IfNoValueTF {#4}
    { #1 \vdash_{\mdart{S}} #2 \Uuparrow #3 \dashv #1' }
    { #1 \vdash_{\mdart{S}} #2 \Uuparrow #3 \dashv #4 }
  } 
}

\DeclareDocumentCommand{\exposuremmd}{ o m m m o } {
  \IfNoValueTF {#1}
  {
    \IfNoValueTF {#5}
    { DS_{#2} \vdash_{\mdart{S}} #2.#3 \Uparrow^{\mu} #4 \dashv DS'_{#2} }
    { DS_{#2} \vdash_{\mdart{S}} #2.#3 \Uparrow^{\mu} #4 \dashv #5 }
  }
  {
    \IfNoValueTF {#5}
    { #1 \vdash_{\mdart{S}} #2.#3 \Uparrow^{\mu} #4 \dashv #1'_{#2} }
    { #1 \vdash_{\mdart{S}} #2.#3 \Uparrow^{\mu} #4 \dashv #5 }
  } 
}

\DeclareDocumentCommand{\imposuremmd}{ o m m m o } {
  \IfNoValueTF {#1}
  {
    \IfNoValueTF {#5}
    { DS_{#2} \vdash_{\mdart{S}} #2.#3 \Downarrow^{\mu} #4 \dashv DS'_{#2} }
    { DS_{#2} \vdash_{\mdart{S}} #2.#3 \Downarrow^{\mu} #4 \dashv #5 }
  }
  {
    \IfNoValueTF {#5}
    { #1 \vdash_{\mdart{S}} #2.#3 \Downarrow^{\mu} #4 \dashv #1'_{#2} }
    { #1 \vdash_{\mdart{S}} #2.#3 \Downarrow^{\mu} #4 \dashv #5 }
  } 
}

\DeclareDocumentCommand{\stareatmdd}{ o m m o } {
  \IfNoValueTF {#1}
  {
    \IfNoValueTF {#4}
    { \Gamma_1 \gg  \{#2\} <:_D \{#3\} \ll \Gamma_2 }
    { \Gamma_1 \gg  \{#2\} <:_D \{#3\} \ll #4 }
  }
  {
    \IfNoValueTF {#4}
    { #1 \gg  \{#2\} <:_D \{#3\} \ll \Gamma_2 }
    { #1 \gg  \{#2\} <:_D \{#3\} \ll #4 }
  } 
}

\DeclareDocumentCommand{\convertible}{o m m m} {
  \IfNoValueTF {#1}
  {
    \Gamma \vdash_{\mdart} #2 : #3 \leadsto #4
  }
  {
    #1 \vdash_{\mdart} #2 : #3 \leadsto #4
  } 
}

\DeclareDocumentCommand{\opesubmd}{ m o } {
  \IfNoValueTF {#2}
  {#1 \subseteq_{\leadsto} #1'}
  {#1 \subseteq_{\leadsto} #2}
}

\DeclareDocumentCommand{\typcheckmd}{ o m m } {
  \IfNoValueTF {#1}
  {\Gamma \vdash_{\mdart{S}} #2 \overleftarrow{:} #3}
  {#1 \vdash_{\mdart{S}} #2 \overleftarrow{:} #3}
}

\DeclareDocumentCommand{\typsynmd}{ o m m m } {
  \IfNoValueTF {#1}
  {\Gamma \vdash_{\mdart{S}} #2 / #3 \overrightarrow{:} #4}
  {#1 \vdash_{\mdart{S}} #2 / #3 \overrightarrow{:} #4}
}

\DeclareDocumentCommand{\expomd}{ o m m } {
  \IfNoValueTF {#1}
  {\Gamma \vdash_{\mdart{S}} #2 \Uparrow #3}
  {#1 \vdash_{\mdart{S}} #2 \Uparrow #3}
}

\DeclareDocumentCommand{\impomd}{ o m m } {
  \IfNoValueTF {#1}
  {\Gamma \vdash_{\mdart{S}} #2 \Downarrow #3}
  {#1 \vdash_{\mdart{S}} #2 \Downarrow #3}
}

\DeclareDocumentCommand{\promomd}{ o m m m } {
  \IfNoValueTF {#1}
  {\Gamma \vdash_{\mdart{S}} #2 \Uparrow_{#3} #4}
  {#1 \vdash_{\mdart{S}} #2 \Uparrow_{#3} #4}
}

\DeclareDocumentCommand{\demomd}{ o m m m } {
  \IfNoValueTF {#1}
  {\Gamma \vdash_{\mdart{S}} #2 \Downarrow_{#3} #4}
  {#1 \vdash_{\mdart{S}} #2 \Downarrow_{#3} #4}
}

\DeclareDocumentCommand{\promdemomd}{ o m m m } {
  \IfNoValueTF {#1}
  {\Gamma \vdash_{\mdart{S}} #2 \Uparrow_{#3}(\Downarrow_{#3}) #4}
  {#1 \vdash_{\mdart{S}} #2 \Uparrow_{#3}(\Downarrow_{#3}) #4}
}

\DeclareDocumentCommand{\dempromomd}{ o m m m } {
  \IfNoValueTF {#1}
  {\Gamma \vdash_{\mdart{S}} #2 \Downarrow_{#3}(\Uparrow_{#3}) #4}
  {#1 \vdash_{\mdart{S}} #2 \Downarrow_{#3}(\Uparrow_{#3}) #4}
}

\DeclareDocumentCommand{\promodmd}{ o m m m } {
  \IfNoValueTF {#1}
  {\Gamma \vdash_{\mdart{SD}} \{#2\} \Uparrow_{#3} \{#4\}}
  {#1 \vdash_{\mdart{SD}} \{#2\} \Uparrow_{#3} \{#4\}}
}

\DeclareDocumentCommand{\demodmd}{ o m m m } {
  \IfNoValueTF {#1}
  {\Gamma \vdash_{\mdart{SD}} \{#2\} \Downarrow_{#3} \{#4\}}
  {#1 \vdash_{\mdart{SD}} \{#2\} \Downarrow_{#3} \{#4\}}
}

\DeclareDocumentCommand{\promdemodmd}{ o m m m } {
  \IfNoValueTF {#1}
  {\Gamma \vdash_{\mdart{SD}} \{#2\} \Uparrow_{#3}(\Downarrow_{#3}) \{#4\}}
  {#1 \vdash_{\mdart{SD}} \{#2\} \Uparrow_{#3}(\Downarrow_{#3}) \{#4\}}
}

\DeclareDocumentCommand{\dempromodmd}{ o m m m } {
  \IfNoValueTF {#1}
  {\Gamma \vdash_{\mdart{SD}} \{#2\} \Downarrow_{#3}(\Uparrow_{#3}) \{#4\}}
  {#1 \vdash_{\mdart{SD}} \{#2\} \Downarrow_{#3}(\Uparrow_{#3}) \{#4\}}
}

\DeclareDocumentCommand{\promommd}{ o m m m } {
  \IfNoValueTF {#1}
  {DS_#3 \vdash_{\mdart{S}} #2 \Uparrow^\mu_{#3} #4}
  {#1 \vdash_{\mdart{S}} #2 \Uparrow^\mu_{#3} #4}
}

\DeclareDocumentCommand{\demommd}{ o m m m } {
  \IfNoValueTF {#1}
  {DS_#3 \vdash_{\mdart{S}} #2 \Downarrow^\mu_{#3} #4}
  {#1 \vdash_{\mdart{S}} #2 \Downarrow^\mu_{#3} #4}
}

\DeclareDocumentCommand{\promdemommd}{ o m m m } {
  \IfNoValueTF {#1}
  {DS_#3 \vdash_{\mdart{S}} #2 \Uparrow^\mu_{#3}(\Downarrow^\mu_{#3}) #4}
  {#1 \vdash_{\mdart{S}} #2 \Uparrow^\mu_{#3}(\Downarrow^\mu_{#3}) #4}
}

\DeclareDocumentCommand{\dempromommd}{ o m m m } {
  \IfNoValueTF {#1}
  {DS_#3 \vdash_{\mdart{S}} #2 \Downarrow^\mu_{#3}(\Uparrow^\mu_{#3}) #4}
  {#1 \vdash_{\mdart{S}} #2 \Downarrow^\mu_{#3}(\Uparrow^\mu_{#3}) #4}
}

\DeclareDocumentCommand{\promomdmd}{ o m m m } {
  \IfNoValueTF {#1}
  {DS_#3 \vdash_{\mdart{SD}} \{#2\} \Uparrow^\mu_{#3} \{#4\}}
  {#1 \vdash_{\mdart{SD}} \{#2\} \Uparrow^\mu_{#3} \{#4\}}
}

\DeclareDocumentCommand{\demomdmd}{ o m m m } {
  \IfNoValueTF {#1}
  {DS_#3 \vdash_{\mdart{SD}} \{#2\} \Downarrow^\mu_{#3} \{#4\}}
  {#1 \vdash_{\mdart{SD}} \{#2\} \Downarrow^\mu_{#3} \{#4\}}
}

\DeclareDocumentCommand{\promdemomdmd}{ o m m m } {
  \IfNoValueTF {#1}
  {DS_#3 \vdash_{\mdart{SD}} \{#2\} \Uparrow^\mu_{#3}(\Downarrow^\mu_{#3}) \{#4\}}
  {#1 \vdash_{\mdart{SD}} \{#2\} \Uparrow^\mu_{#3}(\Downarrow^\mu_{#3}) \{#4\}}
}

\DeclareDocumentCommand{\dempromomdmd}{ o m m m } {
  \IfNoValueTF {#1}
  {DS_#3 \vdash_{\mdart{SD}} \{#2\} \Downarrow^\mu_{#3}(\Uparrow^\mu_{#3}) \{#4\}}
  {#1 \vdash_{\mdart{SD}} \{#2\} \Downarrow^\mu_{#3}(\Uparrow^\mu_{#3}) \{#4\}}
}

\DeclareDocumentCommand{\typingjdot}{ o m m } {
  \IfNoValueTF {#1}
  {\Gamma \vdash_{\jDOT} #2 : #3}
  {#1 \vdash_{\jDOT} #2 : #3}
}

\DeclareDocumentCommand{\typingdjdot}{ o m m m } {
  \IfNoValueTF {#1}
  {\Gamma \vdash_{\jDOT} #3 :_{#2} #4}
  {#1 \vdash_{\jDOT} #3 :_{#2} #4}
}

\DeclareDocumentCommand{\subtypingjdot}{ o m m } {
  \IfNoValueTF {#1}
  {\Gamma \vdash_{\jDOT} #2 <: #3}
  {#1 \vdash_{\jDOT} #2 <: #3}
}

\DeclareDocumentCommand{\ucjdot}{ o m m o } {
  \IfNoValueTF {#1}
  {
    \IfNoValueTF {#4}
    { \Gamma \vdash_{\jDOT{S}} #2 \nearrow #3 \dashv \Gamma' }
    { \Gamma \vdash_{\jDOT{S}} #2 \nearrow #3 \dashv #4 }
  }
  {
    \IfNoValueTF {#4}
    { #1 \vdash_{\jDOT{S}} #2 \nearrow #3 \dashv #1' }
    { #1 \vdash_{\jDOT{S}} #2 \nearrow #3 \dashv #4 }
  } 
}

\DeclareDocumentCommand{\dcjdot}{ o m m o } {
  \IfNoValueTF {#1}
  {
    \IfNoValueTF {#4}
    { \Gamma \vdash_{\jDOT{S}} #2 \searrow #3 \dashv \Gamma' }
    { \Gamma \vdash_{\jDOT{S}} #2 \searrow #3 \dashv #4 }
  }
  {
    \IfNoValueTF {#4}
    { #1 \vdash_{\jDOT{S}} #2 \searrow #3 \dashv #1' }
    { #1 \vdash_{\jDOT{S}} #2 \searrow #3 \dashv #4 }
  } 
}

\DeclareDocumentCommand{\udcjdot}{ o m m o } {
  \IfNoValueTF {#1}
  {
    \IfNoValueTF {#4}
    { \Gamma \vdash_{\jDOT{S}} #2 \nearrow(\searrow) #3 \dashv \Gamma' }
    { \Gamma \vdash_{\jDOT{S}} #2 \nearrow(\searrow) #3 \dashv #4 }
  }
  {
    \IfNoValueTF {#4}
    { #1 \vdash_{\jDOT{S}} #2 \nearrow(\searrow) #3 \dashv #1' }
    { #1 \vdash_{\jDOT{S}} #2 \nearrow(\searrow) #3 \dashv #4 }
  } 
}

\DeclareDocumentCommand{\revealingjdot}{ o m m o } {
  \IfNoValueTF {#1}
  {
    \IfNoValueTF {#4}
    { \Gamma \vdash_{\jDOT{S}} #2 \Uuparrow #3 \dashv \Gamma' }
    { \Gamma \vdash_{\jDOT{S}} #2 \Uuparrow #3 \dashv #4 }
  }
  {
    \IfNoValueTF {#4}
    { #1 \vdash_{\jDOT{S}} #2 \Uuparrow #3 \dashv #1' }
    { #1 \vdash_{\jDOT{S}} #2 \Uuparrow #3 \dashv #4 }
  } 
}

\DeclareDocumentCommand{\promojdot}{ o m m m } {
  \IfNoValueTF {#1}
  {\Gamma \vdash_{\jDOT{S}} #2 \Uparrow_{#3} #4}
  {#1 \vdash_{\jDOT{S}} #2 \Uparrow_{#3} #4}
}

\DeclareDocumentCommand{\demojdot}{ o m m m } {
  \IfNoValueTF {#1}
  {\Gamma \vdash_{\jDOT{S}} #2 \Downarrow_{#3} #4}
  {#1 \vdash_{\jDOT{S}} #2 \Downarrow_{#3} #4}
}

\DeclareDocumentCommand{\promdemojdot}{ o m m m } {
  \IfNoValueTF {#1}
  {\Gamma \vdash_{\jDOT{S}} #2 \Uparrow_{#3}(\Downarrow_{#3}) #4}
  {#1 \vdash_{\jDOT{S}} #2 \Uparrow_{#3}(\Downarrow_{#3}) #4}
}

\DeclareDocumentCommand{\dempromojdot}{ o m m m } {
  \IfNoValueTF {#1}
  {\Gamma \vdash_{\jDOT{S}} #2 \Downarrow_{#3}(\Uparrow_{#3}) #4}
  {#1 \vdash_{\jDOT{S}} #2 \Downarrow_{#3}(\Uparrow_{#3}) #4}
}

\DeclareDocumentCommand{\convertiblejdot}{ o m m m o } {
  \IfNoValueTF {#1}
  {
    \IfNoValueTF {#5}
    { \Gamma_1 \rhd #2 \xRightarrow{#3} #4 \lhd \Gamma_2 }
    { \Gamma_1 \rhd #2 \xRightarrow{#3} #4 \lhd \ll #5 }
  }
  {
    \IfNoValueTF {#5}
    { #1 \rhd #2 \xRightarrow{#3} #4 \lhd \Gamma_2 }
    { #1 \rhd #2 \xRightarrow{#3} #4 \lhd #5 }
  } 
}

\DeclareDocumentCommand{\typcheckjdot}{ o m m } {
  \IfNoValueTF {#1}
  {\Gamma \vdash_{\jDOT{S}} #2 \overleftarrow{:} #3}
  {#1 \vdash_{\jDOT{S}} #2 \overleftarrow{:} #3}
}

\DeclareDocumentCommand{\typsynjdot}{ o m m m } {
  \IfNoValueTF {#1}
  {\Gamma \vdash_{\jDOT{S}} #2 / #3 \overrightarrow{:} #4}
  {#1 \vdash_{\jDOT{S}} #2 / #3 \overrightarrow{:} #4}
}

\DeclareDocumentCommand{\synthesizer}{ o m m m m } {
  \IfNoValueTF {#1}
  {\Gamma \rhd_{#2} #3 / #4 \overrightarrow{:} #5}
  {#1 \rhd_{#2} #3 / #4 \overrightarrow{:} #5}
}

\DeclareDocumentCommand{\typcheckdjdot}{ o m m m } {
  \IfNoValueTF {#1}
  {\Gamma \vdash_{\jDOT{S}} #3 \overleftarrow{:_{#2}} #4}
  {#1  \vdash_{\jDOT{S}} #3 \overleftarrow{:_{#2}} #4}
}

\DeclareDocumentCommand{\subtypingjdotn}{ o m m m } {
  \IfNoValueTF {#1}
  {\Gamma \vdash_{\jDOT} #2 <:_{#3} #4}
  {#1 \vdash_{\jDOT} #2 <:_{#3} #4}
}

\DeclareDocumentCommand{\subtypingfmL}{ m o m m } {
  \IfNoValueTF {#2}
  {[#1] \Gamma \vdash_{\fsub^-} #3 <: #4}
  {[#1] #2 \vdash_{\fsub^-} #3 <: #4}
}

\DeclareDocumentCommand{\subtypingfms}{ o m m m } {
  \IfNoValueTF {#1}
  {\Gamma \vdash[#2]_{\fsub^-} #3 <: #4}
  {#1 \vdash[#2]_{\fsub^-} #3 <: #4}
}

\begin{document}

\title{Undecidability of $\dsub$ and Its Decidable Fragments}         


\author{Jason Hu}
\affiliation{
  \department{Cheriton School of Computer Science}              
  \institution{University of Waterloo}            
  \streetaddress{200 University Avenue West}
  \city{Waterloo}
  \state{Ontario}
  \postcode{N2l 3G1}
  \country{Canada}                    
}
\email{zs2hu@uwaterloo.ca}          

\author{Ond\v{r}ej Lhot\'{a}k}
\orcid{nnnn-nnnn-nnnn-nnnn}             
\affiliation{
  \position{Associate Professor}
  \department{Cheriton School of Computer Science}              
  \institution{University of Waterloo}            
  \streetaddress{200 University Avenue West}
  \city{Waterloo}
  \state{Ontario}
  \postcode{N2l 3G1}
  \country{Canada}                    
}
\email{olhotak@uwaterloo.ca}         

\begin{abstract}
  Dependent Object Types (\DOT) is a calculus with path dependent types, intersection
types, and object self-references, which serves as the core calculus of Scala 3.
Although the calculus has been proven sound, it remains open whether
type checking in \DOT is decidable.
In this paper, we establish
undecidability proofs of type checking and subtyping of $\dsub$, a syntactic subset of
\DOT. It turns out that even for $\dsub$, undecidability is surprisingly difficult to show,
as evidenced by counterexamples for past attempts.
To prove undecidability, we discover an equivalent definition of
the $\dsub$ subtyping rules in normal form. Besides being easier to reason about, this
definition makes the phenomenon of bad bounds explicit as a single inference rule.
After removing this rule, we
discover two decidable fragments of $\dsub$ subtyping and identify algorithms to decide them.
We prove soundness and completeness of the algorithms with respect to the fragments, and
we prove that the algorithms terminate.
Our proofs are mechanized in a combination of Coq and Agda.

\end{abstract}



\maketitle

\section{Introduction}

The Dependent Object Types (\DOT) calculus has received attention as a
model for the Scala type system~%
\citep[etc.]{wadlerfest-dot,oopsla-dot,simple-sound-proof}.
The calculus features objects with abstract type members with upper and
lower bounds, and path-dependent types to select those type members.
It also supports object self-references, intersection types, and dependent function types.

To implement any type system in a compiler requires a type checking \emph{algorithm}.
If type checking is undecidable, a compiler writer needs either at least a semi-algorithm
or an algorithm for a decidable variant of the type system.

Type checking \DOT has been conjectured to be undecidable because bounded quantification
is undecidable in $\fsub$~\citep{fsub-undec}. However, such informal reasoning about
\DOT can understandably be incorrect, as we show with a simple example in
\Cref{ssec:ipf}. Formally determining decidability of \DOT turns out to
be surprisingly challenging. It is challenging even for $\dsub$, a restriction
of \DOT that removes self-references and intersection types, leaving type
members and path-dependent types that select them~%
\citep{wadlerfest-dot,defint}. 
In this paper, our focus is entirely on decidability of $\dsub$ and its variants.

\newcommand{\stepone}{\textcircled{\small 1}\xspace}
\newcommand{\steptwo}{\textcircled{\small 2}\xspace}
\newcommand{\stepthree}{\textcircled{\small 3}\xspace}

A general technique to prove a decision problem $P$ undecidable is
\emph{reduction} from a known undecidable problem $Q$. This requires
\stepone defining a mapping $M$ from instances of $Q$ to
instances of $P$ and proving that $p$ is yes-instance of $P$
if \steptwo and only if \stepthree $q$ is a yes-instance of $Q$.
\citet{wadlerfest-dot} does \stepone and \steptwo for a reduction from
$\fsub$ to $\dsub$. However, in \Cref{ssec:ipf}, we identify a counterexample
to \stepthree. This means that the proposed mapping \stepone cannot be used to
prove $\dsub$ undecidable.

Based on the counterexample, we define $\fsubm$, an undecidable fragment
of $\fsub$ that is better suited for reduction to $\dsub$. However,
reduction is still thwarted by subtyping transitivity, which is posed
as an explicit inference rule in $\dsub$. In $\dsub$, all reasoning about any subtyping
relationship $S <: U$ must consider the possibility that it arose due
to transitivity $S <: T <: U$ involving some arbitrary and unknown type $T$.

In previous work on \DOT and $\dsub$, a recurring challenge
has been the concept of \emph{bad bounds}. In the presence of a type
member declaration $x: \{A:S..U\}$ with upper and lower bounds, the
defining subtyping relationships $S <: x.A$ and $x.A <: U$ conspire with
transitivity to induce the possibly unexpected and undesirable subtyping
relationship $S <: U$ between the bounds.

For $\fsub$, there is a normal form of the subtyping
rules that achieves transitivity without an explicit rule~%
\citep{10.1007/3-540-52590-4_45,fsub-undec}.
We discover an analogous normal form for $\dsub$ in \Cref{ssec:nf}.
In particular, we show that to achieve transitivity in $\dsub$ normal   
form, it is both necessary and sufficient to express the bad bounds
concept as an explicit rule (\infref{dsub:bb}), and add it
to the obvious fundamental rules that define the meaning of each form of type.
$\dsub$ normal form turns out to have convenient properties and becomes the
core concept underlying all of our developments.

We prove undecidability of $\dsub$ by a reduction from $\fsubm$ to
$\dsub$ normal form.

In $\dsub$ normal form, undecidability is crisply characterized by
two specific subtyping rules. The first is the \infref{dsub:all} rule
that compares function types, which is well known from $\fsub$ as the root
cause of its undecidability. In $\fsub$, this rule can be restricted to
a kernel version that applies only to functions with equal parameter
types to make the resulting \emph{kernel} $\fsub$ decidable.
The second is the \infref{dsub:bb} rule that models bad bounds.
If the \infref{dsub:bb} rule is removed from $\dsub$ and the \infref{dsub:all} rule
is replaced with the kernel version, the resulting kernel $\dsub$
becomes decidable.

Moreover, we show that kernel $\dsub$ is exactly
fragment of (full) $\dsub$ that can be typed by the partial
typing algorithm of~%
\citet{abel-algorithmic}.
Nieto identified a counterexample demonstrating that the subtyping
relation implemented by the Scala compiler violates transitivity.
The violation corresponds directly to the \infref{dsub:bb} rule of
kernel $\dsub$. The implementation of subtyping in the compiler does not implement this
rule. This observation motivates dropping this problematic rule
from practical, decidable variants of $\dsub$ normal form and
\DOT (when a normal form for \DOT is found).

The kernel restriction of the \infref{dsub:all} rule seriously
limits expressiveness in $\dsub$ because it prevents comparison
between parameter types of functions. This disables the case in which the parameter types are type aliases of each
other. For example, in the scope of a type member declaration
$x: \{A: T..T\}$, the types $x.A$ and $T$ should be considered
equivalent. To address this limitation, we define a \emph{strong
kernel} variant of the \infref{dsub:all} rule that allows comparison between parameter
types. The expressiveness of strong kernel $\dsub$ is strictly between
kernel $\dsub$ and full $\dsub$, but unlike full $\dsub$, strong kernel $\dsub$
is decidable. Finally, we provide stare-at subtyping, an algorithm to
decide subtyping in strong kernel $\dsub$.

To summarize, our contributions are:
\begin{enumerate}
    \item a counterexample to the previously proposed reduction from $\fsub$ to $\dsub$,
    \item $\dsub$ normal form and its equivalence to $\dsub$,
    \item undecidability of $\dsub$ by reduction from $\fsubm$,
    \item equivalence of kernel $\dsub$ and the fragment of $\dsub$ typeable by Nieto's algorithm,
    \item strong kernel $\dsub$, and
    \item the stare-at algorithm for typing strong kernel $\dsub$.
\end{enumerate}
We have verified the proofs of our lemmas and theorems in a combination of Coq and Agda.
The formalization is included as supplementary material and will be submitted as an artifact.

The properties of the variants of $\dsub$ are summarized in \Cref{tab:summary}. 

\begin{table}[]
  \begin{tabular}{|l|l|l|l|}
    \hline
    Name                            & \infref{dsub:all} rule & \infref{dsub:bb} rule & Decidability                               \\ \hline
    $\dsub$ and $\dsub$ normal form & full \infref{dsub:all}               & $\checkmark$     & undecidable (\Cref{ssec:nf})        \\
                                    & full \infref{dsub:all}               & $\times$         & undecidable (\Cref{ssec:nf})        \\
    Strong kernel $\dsub$           & \infref{dsub:sk:all}   & $\times$         & decidable by Stare-at subtyping (\Cref{ssec:stareat}) \\
    Kernel $\dsub$                  & kernel \infref{dsub:k:all}           & $\times$         & decidable by Step subtyping (\Cref{ssec:step})        \\ \hhline{|=|=|=|=|}
                                    & \infref{dsub:k:all} or \infref{dsub:sk:all} & $\checkmark$     & unknown                                 \\\hline
  \end{tabular}
%
%
%
%
  \caption{Summary of $\dsub$ variants}
  \label{tab:summary}
\end{table}

\section{Preliminaries}

We adopt the following conventions throughout the paper.

  Throughout this paper, we consider two types or terms the same if they are
  equivalent up to $\alpha$-conversion~\cite{barendregt1984lambda}. We use subscripts
  to emphasize free occurrences of a variable. For example, $T_x$ means $x$ may have
  free occurrences in $T$. Additionally, we assume $\alpha$-conversion happens
  automatically. That is, when $T_y$ appears later, all corresponding free $x$'s in
  $T$ are substituted by $y$.

  We use semicolons ($;$) to denote context concatenation instead of commas ($,$).


\begin{definition}
  A type $T$ is closed w.r.t. a context $\Gamma$, if $fv(T) \subseteq dom(\Gamma)$.
\end{definition}

\begin{definition}
  Well-formedness of a context is inductively defined as follows.
  \begin{enumerate}
  \item The empty context $\bigcdot$ is well-formed.
  \item If $\Gamma$ is well-formed, $T$ is closed w.r.t. $\Gamma$ and $x \notin
    dom(\Gamma)$, then $\Gamma ; x : T$ is well-formed.
  \end{enumerate}
\end{definition}

Unless explicitly mentioned, all lemmas and theorems require and preserve that
types are closed and contexts are well-formed. This is proven explicitly in the
mechanized proofs.

\section{Definitions of $\fsub$ and $\dsub$}\label{ssec:defs}

$\fsub$ is introduced by \citet{Cardelli:1985:UTD:6041.6042} as the core calculus of
the Fun language, which extends system $F$ with upper bounded quantification.

\begin{figure}
  \vspace{-20pt}
  \begin{multicols}{2}
    \begin{align*}
      X,Y,Z & & \tag*{\textbf{Type variable}} \\
      S, T, U ::= & & \tag*{\textbf{Type}} \\
                  & \top & \tag*{top type} \\
                  & X & \tag*{type variable} \\
                  & S \to U & \tag*{function} \\
                  & \forall X <: S . U_X & \tag*{universal type}
    \end{align*}
    
    \begin{mathpar}
      \inferrule*[right=\infnlabel{F-All}{fsub:all}]
      {\subtypingf{S'}{S} \\ \subtypingf[\Gamma; X <: S']{U}{U'}}
      {\subtypingf{\forall X<:S.U}{\forall X <: S' .U'}}
    \end{mathpar}
    
    \begin{mathpar}
      \inferrule*[right=\infnlabel{F-Top}{fsub:top}]
      { }{\subtypingf{T}\top}

      \inferrule*[right=\infnlabel{F-Refl}{fsub:refl}]
      { }{\subtypingf T T}
      
      \inferrule*[right=\infnlabel{F-Tvar}{fsub:tvar}]
      {X <: T \in \Gamma}
      {\subtypingf X T}

      \inferrule*[right=\infnlabel{F-Fun}{fsub:fun}]
      {\subtypingf{S'}{S} \\ \subtypingf{U}{U'}}
      {\subtypingf{S \to U}{S' \to U'}}
      
      \inferrule*[right=\infnlabel{F-Trans}{fsub:trans}]
      {\subtypingf S T \\ \subtypingf T U}
      {\subtypingf S U}
    \end{mathpar}
  \end{multicols}
  
  \caption{Definition of subtyping in $\fsub$~\cite[Figure
    26-2]{tapl}}\label{fig:fsub_def}
\end{figure}

\begin{definition}
  $\fsub$ is defined in \Cref{fig:fsub_def}.
\end{definition}

Universal types in $\fsub$ combine polymorphism and subtyping. Universal types can be
compared by the \infref{fsub:all} rule. The \infref{fsub:trans} rule indicates that
the system has transitivity. It turns out that $\fsub$ can be defined in a way such
that transitivity does not appear as an inference rule but rather a provable property.

\begin{figure}
  \begin{mathpar}
    \inferrule*[right=\infref{fsub:top}]
    { }{\subtypingf{T}\top}

    \colorbox{light-gray}{
      $\inferrule*[right=\infnlabel{F-VarRefl}{fsubnf:varrefl}]
      { }{\subtypingf X X}$
    }
    
    \colorbox{light-gray}{
      $\inferrule*[right=\infnlabel{F-Tvar'}{fsubnf:tvar}]
      {X <: T \in \Gamma \\ \subtypingf T U}
      {\subtypingf X U}$
    }

    \inferrule*[right=\infref{fsub:fun}]
    {\subtypingf{S'}{S} \\ \subtypingf{U}{U'}}
    {\subtypingf{S \to U}{S' \to U'}}

    \inferrule*[right=\infref{fsub:all}]
    {\subtypingf{S'}{S} \\ \subtypingf[\Gamma; X <: S']{U}{U'}}
    {\subtypingf{\forall X<:S.U}{\forall X <: S' .U'}}
  \end{mathpar}

  \caption{Definition of $\fsub$ normal form}\label{fig:fsub_nf_def}
\end{figure}

\begin{definition}
  $\fsub$ normal form is defined in \Cref{fig:fsub_nf_def}. The different rules are
  shaded.
\end{definition}

We call this alternative definition ``normal form'', following the convention
in~\citet{fsub-undec}. Both definitions are equivalent:

\begin{theorem}\label{thm:fsub-nf-equiv}\citep{10.1007/3-540-52590-4_45}
  $\fsub$ subtyping is equivalent to $\fsub$ normal form. Namely $\subtypingf S U$
  holds in non-normal form, iff it holds in normal form.
\end{theorem}

$\dsub$ is a richer calculus than $\fsub$. It adds a form of dependent types, called
path types, each of which has both upper bounds and lower bounds, so it is more
general than $\fsub$.

\begin{figure}
  \vspace{-20pt}
  \begin{multicols}{2}
    \begin{align*}
      x,y,z & & \tag*{\textbf{Variable}} \\
      S, T, U ::= & & \tag*{\textbf{Type}} \\
                  & \top & \tag*{top type} \\
                  & \bot & \tag*{bottom type} \\
                  & \{A:S..U\} & \tag*{type declaration} \\
                  & x.A & \tag*{path type} \\
                  & \forall(x : S) U_x & \tag*{function}
    \end{align*}

    \begin{align*}
      v ::= & & \tag*{\textbf{Value}} \\
            & \{A = T\} & \tag*{type tag}\\
            & \lambda(x : T)t_x & \tag*{lambda} \\
      s,t,u ::= & & \tag*{\textbf{Term}} \\
            & x & \tag*{variable} \\
            & v & \tag*{value} \\
            & x\ y & \tag*{application} \\
            & \letin{x}{t}{u_x} & \tag*{let binding}
    \end{align*}
  \end{multicols}
  \textbf{Type Assignment}  
  \begin{mathpar}
    \inferrule*[right=\infnlabel{Var}{dsub:var}]{ }{\typingd{x}{\Gamma(x)}}

    \inferrule*[right=\infnlabel{Sub}{dsub:sub}]
    {\typingd{t}{S} \\ \subtypingd{S}{U}}
    {\typingd{t}{U}}

    \inferrule*[right=\infnlabel{All-I}{dsub:all-i}]
    {\typingd[\Gamma; x : S]{t}{U_x}}
    {\typingd{\lambda(x : S)t}{\forall(x : S)U_x}}

    \inferrule*[right=\infnlabel{All-E}{dsub:all-e}]
    {\typingd{x}{\forall(z: S)U_z} \\\\ \typingd{y}{S}}
    {\typingd{x\ y}{U_y}}

    \inferrule*[right=\infnlabel{Typ-I}{dsub:typ-i}]
    { } {\typingd{\{A = T \}}{\{A : T .. T\}}}

    \inferrule*[right=\infnlabel{Let}{dsub:let}]
    {\typingd{t}{S} \\ x \notin fv(U) \\\\ \typingd[\Gamma; x : S]{u}{U}}
    {\typingd{{\letin{x}{t}{u}}}U}
  \end{mathpar}
  \textbf{Subtyping}
  \begin{mathpar}
    \inferrule*[right=\infnlabel{Top}{dsub:top}]
    { }{\subtypingd{T}\top}

    \inferrule*[right=\infnlabel{Bot}{dsub:bot}]
    { }{\subtypingd \bot T}

    \inferrule*[right=\infnlabel{Refl}{dsub:refl}]
    { }{\subtypingd T T}

    \inferrule*[right=\infnlabel{Bnd}{dsub:bnd}]
    {\subtypingd{S_2}{S_1} \\ \subtypingd{U_1}{U_2}}
    {\subtypingd{\{ A: S_1 .. U_1 \}}{\{A:S_2 .. U_2\}}}
    
    \inferrule*[right=\infnlabel{All}{dsub:all}]
    {\subtypingd{S_2}{S_1} \\ \subtypingd[\Gamma; x : S_2]{U_1}{U_2}}
    {\subtypingd{\forall(x: S_1)U_1}{\forall(x: S_2)U_2}}

    \inferrule*[right=\infnlabel{Sel1}{dsub:sel1}]
    {\typingd{x}{\{A: S .. U\}}}
    {\subtypingd{S}{x.A}}
    
    \inferrule*[right=\infnlabel{Sel2}{dsub:sel2}]
    {\typingd{x}{\{A: S .. U\}}}
    {\subtypingd{x.A}{U}}
    
    \inferrule*[right=\infnlabel{Trans}{dsub:trans}]
    {\subtypingd S T \\ \subtypingd T U}
    {\subtypingd S U}
  \end{mathpar}
  
  \caption{Definition of $\dsub$~\citep{wadlerfest-dot}}\label{fig:dsub_def}
\end{figure}

\begin{definition}
  $\dsub$ is defined in \Cref{fig:dsub_def}.
\end{definition}

$\dsub$ has the following types: the top type $\top$, the bottom type $\bot$, type
declarations, path types, and dependent function types. In $\dsub$, a path type has
the form $x.A$ where the type label $A$ is fixed. That is, in $\dsub$, there is only
one type label and it is $A$. A term in $\dsub$ can be a variable, a type tag, a
lambda abstraction, an application, or a let binding.

In the typing rules, the \infref{dsub:var}, \infref{dsub:sub} and \infref{dsub:let}
rules are standard. The \infref{dsub:all-i} rule says a lambda is typed by pushing its
declared parameter type to the context. Note that the return type is allowed to depend
on the parameter, which makes the system dependently typed. The \infref{dsub:all-e}
rule types a function application. Since $U$ may depend on its parameter, the overall
type may refer to $y$.  The \infref{dsub:typ-i} rule assigns a type declaration with
equal bounds to a type tag.

In the subtyping rules, the \infref{dsub:top}, \infref{dsub:bot}, \infref{dsub:refl}
and \infref{dsub:trans} rules are standard. In the \infref{dsub:bnd} rule, type
declarations are compared by comparing their corresponding components. Notice that the
lower bounds are in contravariant position and hence they are compared in reversed
order. Similarly, the \infref{dsub:all} rule also compares parameter types in reversed
order. The return types are compared with the context extended with $S_2$. The
\infref{dsub:sel1} and \infref{dsub:sel2} rules are used to access the bounds of a
path type.

Notice that the typing and subtyping rules in $\dsub$ are mutually dependent. This is
because the \infref{dsub:sub} rule uses subtyping and the \infref{dsub:sel1} and
\infref{dsub:sel2} rules use typing in their premises. This mutual dependency makes
$\dsub$ harder to reason about. Nonetheless, this mutual dependency can be eliminated due to
the following lemma.

\begin{lemma}\label{lem:dsub-unravel} (unravelling of $\dsub$ subtyping)
  \infref{dsub:sel1} and \infref{dsub:sel2} can be changed to the following rules, and
  the resulting subtyping relation is equivalent to the original one.
  \begin{mathpar}
    \inferrule*[right=\infnlabel{Sel1'}{dsub:sel1p}]
    {\subtypingd{\Gamma(x)}{\{A: S .. \top\}}}
    {\subtypingd{S}{x.A}}

    \inferrule*[right=\infnlabel{Sel2'}{dsub:sel2p}]
    {\subtypingd{\Gamma(x)}{\{A: \bot .. U\}}}
    {\subtypingd{x.A}{U}}
  \end{mathpar}
\end{lemma}

\begin{proof}
  This follows directly from the fact that the only typing rules that apply to
  variables are the \infref{dsub:var} and \infref{dsub:sub} rules.
\end{proof}

This new definition of subtyping with the \infref{dsub:sel1p} and
\infref{dsub:sel2p} rules no longer depends on typing. We will use this
definition in the rest of the paper.

\section{Undecidability of $\dsub$ (sub)typing}

\subsection{Definition of Undecidability}\label{ssec:undec-def}

A common method for proving a decision problem undecidable is by \emph{reduction}
from some other known undecidable problem.

\begin{definition} \citep[Definition
  12.1a]{martin2010introduction} \label{def:reduction} If $Q$ and $P$ are decision
  problems, we say $Q$ is reducible to $P$ ($Q \le P$) if there is an algorithmic
  procedure $F$ that allows us, given an arbitrary instance $I_1$ of $Q$, to find an
  instance $F(I_1)$ of $P$ so that for every $I_1$, $I_1$ is a yes-instance of $Q$
  if and only if $F(I_1)$ is a yes-instance of $P$.
\end{definition}

Notice that reducibility requires an if and only if proof: for our choice of
$F$, we must show that $I_1$ is a yes-instance of $Q$ \emph{if and only if}
$F(I_1)$ is a yes-instance of $P$.

Reduction can be understood intuitively as an adversarial game.
Consider a
  target decision problem $P$. Merlin is a wizard who claims to have access to true
  magic, and therefore be able to decide $P$. He is so confident that he would also
  offer a complete proof accompanying each yes answer he gives.
  Sherlock is a skeptical detective. He questions the Merlin's ability, and comes up
  with the following scheme in order to disprove Merlin's claim.

  
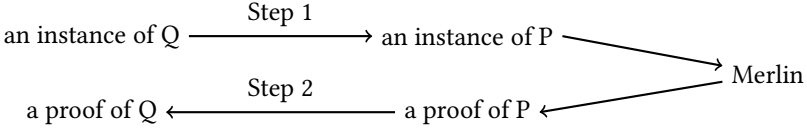
\begin{centering}

    \begin{tikzpicture}
      \node at (0, 0) (Q) {an instance of Q};
      \node at (5, 0) (P) {an instance of P};
      \node at (9, -0.5) (M) {Merlin};
      \node at (5, -1) (P') {a proof of P};
      \node at (0, -1) (Q') {a proof of Q};

      \path[draw, ->, thick] (Q) -- (P) node[midway, above]{Step 1};
      \path [draw, ->, thick] (P.east) -- (M);
      \path [draw, ->, thick] (M) -- (P'.east);
      \path [draw, ->, thick] (P') -- (Q') node[midway, above]{Step 2};
    \end{tikzpicture}

  \end{centering}

  Sherlock selects some undecidable problem $Q$. As Step~1, Sherlock devises a mapping
  from instances of $Q$ to instances of $P$ that preserves yes-instances: every yes-instance
  of $Q$ maps to some yes-instance of $P$. As Step~2, Sherlock devises a mapping from
  (yes-)proofs of $P$ to (yes-)proofs of $Q$. Then, if Merlin could really decide $P$, then Sherlock
  could use this setup to decide $Q$, which is impossible. For any instance of $Q$,
  Sherlock would map it to an instance of $P$ and give it to Merlin to decide. If the
  instance of $P$ is a no-instance, then so was the instance of $Q$. If the instance
  of $P$ is a yes-instance, then Sherlock could map Merlin's proof into a proof that
  the instance of $Q$ is also a yes-instance. 
  $P$ is undecidable if and only if Sherlock achieves both steps and therefore proves Merlin is
  wrong.


\subsection{The Partial Undecidability Proof of \citet{wadlerfest-dot}}\label{ssec:ipf}

Subtyping in $\fsub$ is known to be undecidable~\citep{fsub-undec}.
\citet{wadlerfest-dot} defined the following total mappings from types and contexts
in $\fsub$ to types and contexts in $\dsub$:

\begin{definition}\label{def:mapping}\citep{wadlerfest-dot} The mappings $\inttyp{\cdot}$ and $\intctx{\cdot}$ are defined as follows:
  \begin{align*}
    \inttyp{\top} &= \top \\
    \inttyp{X} &= x_X.A \\
    \inttyp{S \to U} &= \forall(x : \inttyp{S}) \inttyp{U} \tag{function case}\label{eq:fsub:func}
    \\
    \inttyp{\forall X <: S . U} &= \forall(x_X : \{A: \bot .. \inttyp{S}\}) \inttyp{U}
  \end{align*}
  \begin{align*}
    \intctx{\bigcdot} &= \bigcdot \\
    \intctx{\Gamma ; X <: T} &= \intctx{\Gamma} ; x_X : \{A: \bot .. \inttyp{T} \}
  \end{align*}
\end{definition}
In the mapping, a correspondence between type variables in $\fsub$
and variables in $\dsub$ is assumed, as indicated by the notation $x_X$.
\citeauthor{wadlerfest-dot} also proved that given a yes-instance of subtyping in
$\fsub$, its image under the mapping is also a yes-instance of subtyping in $\dsub$:

\begin{theorem}\citep[Theorem 1]{wadlerfest-dot}\label{thm:fsub2dsub}
  If $\subtypingf S U$, then $\subtypingd[\intctx{\Gamma}]{\inttyp{S}}{\inttyp{U}}$. 
\end{theorem}

According to \Cref{def:reduction}, to show subtyping in $\dsub$ undecidable,
it remains to show the other direction:

\begin{conjecture}\label{con:dsub2fsub}
  If $\subtypingd[\intctx{\Gamma}]{\inttyp{S}}{\inttyp{U}}$, then $\subtypingf S U$. 
\end{conjecture}

To see why this step is essential, consider what would happen if we defined a new
calculus $\dsub^+$ by extending $\dsub$ subtyping with a rule that makes every type
$S$ a subtype of every type $U$:

%

\begin{mathpar}
  \inferrule*[right=Trivial]{ }{\subtypingdp S U}
\end{mathpar}
The mapping in \Cref{def:mapping} and \Cref{thm:fsub2dsub} continue to hold
even for $\dsub^+$. But subtyping in $\dsub^+$ is obviously decidable because
every instance is a yes-instance. If \Cref{thm:fsub2dsub} were sufficient to
prove undecidability of $\dsub$, then it would also be sufficient to ``prove''
undecidability of the obviosuly decidable $\dsub^+$. Thus, \Cref{con:dsub2fsub}
is \emph{essential} to complete the proof of undecidability of $\dsub$ subtyping.

%
%

Unfortunately, \Cref{con:dsub2fsub} is \emph{false}. As a counterexample, consider the following subtyping query in $\fsub$:
\begin{align*}
  \subtypingf[]{\top \to \top}{\forall X <: \top. \top}
\end{align*}
This subtyping relationship is false: in $\fsub$, function types and universal types are not related by subtyping. The image of this subtyping relationship under the mapping is:
\begin{align*}
  {\subtypingd[]{\forall(x : \top). \top}{\forall(x_X : \{A : \bot .. \top\}).\top}}
\end{align*}
In $\dsub$, this subtyping relationship is true, 
%
%
%
as witnessed by the
following derivation tree.
\begin{mathpar}
  \inferrule*[right=\infref{dsub:all}]
  {\inferrule*[right=\infref{dsub:top}]{ }{\subtypingd[]{\{A: \bot .. \top\}} \top} \\
    \inferrule*[right=\infref{dsub:refl}]{ }{\subtypingd[x : \{A: \bot .. \top\}] \top \top}}
  {\subtypingd[]{\forall(x : \top). \top}{\forall(x_X : \{A : \bot .. \top\}).\top}}
\end{mathpar}

%

The counterexample shows that the mapping in \Cref{def:mapping} \emph{cannot} be
used to prove undecidability of $\dsub$ subtyping.

%

\subsection{$\fsub^-$}\label{ssec:fsubm}


The counterexample suggests that the problem with the mapping is that it
permits interference between function types and universal types in $\fsub$
because it maps both of them to dependent function types in $\dsub$.
Reviewing \citet{fsub-undec}, we
notice that the undecidability proof of $\fsub$ does not make use of function types.
Therefore, we can remove function types from $\fsub$ to obtain a simpler
calculus that is better suited for undecidability reductions.

\begin{definition}\label{def:fsubm}
  $\fsub^-$ is obtained from $\fsub$ defined in \Cref{fig:fsub_nf_def} by removing function
  types ($\to$) and the \infref{fsub:fun} rule.
\end{definition}

\begin{theorem}\label{thm:fsubminus-undec}
  $\fsub^-$ subtyping is undecidable. 
\end{theorem}
\begin{proof}
  The $\fsub$ undecidability proof of \citet{fsub-undec} does not depend on function types.
\end{proof}

The mappings from \Cref{def:mapping} can be applied to types and contexts in
$\fsubm$. The \ref{eq:fsub:func} can be removed from the mapping since $\fsubm$
does not have function types.


\subsection{Bad Bounds}\label{ssec:attempt}
Since $\fsubm$ invalidates the counterexample to \Cref{con:dsub2fsub},
we can attempt to prove the conjecture for $\fsubm$. When we try to
invert the premise of the conjecture, $\subtypingd[\intctx{\Gamma}]{\inttyp{S}}{\inttyp{U}}$, the first problem we encounter are \emph{bad bounds}.
The pattern of bad bounds is discussed
in~\citet{simple-sound-proof,oopsla-dot}. Bad bounds are an unintended consequence of the
combination of the \infref{dsub:sel1p}, \infref{dsub:sel2p} and \infref{dsub:trans} rules. In certain typing contexts, bad bounds make it possible to prove
subtyping between \emph{any} types $S$ and $U$. Consider the following
derivation tree:
\begin{align*}
  &\text{assume }\Gamma(x) = \{A : S .. U \} \\
  &\inferrule*[right=\infref{dsub:trans}]
  {\inferrule*[right=\infref{dsub:sel1p}]
    {\inferrule*[right=\infref{dsub:refl}]
    { }
    {\typingd{\{A : S .. U \}}{\{A: S .. U\}}}}
    {\subtypingd{S}{x.A}} \\
  \inferrule*[right=\infref{dsub:sel2p}]
  {\text{same as left}}
  {\subtypingd{x.A}{U}}}
  {\subtypingd{S}{U}}
\end{align*}

This derivation uses transitivity to connect the lower and upper bounds of the path
type $x.A$. The types $S$ and $U$ can be any types at all, as long as they appear in the type of $x$ in the typing context $\Gamma$.

In $\fsub$, on the other hand, it is easy to show that, for example, a supertype of
$\top$ must be $\top$. Properties like this are called \emph{inversion properties}.
These properties do not hold in general in $\dsub$ due to bad bounds.
Fortunately, we can prove similar properties in $\dsub$ if we restrict the
typing context $\Gamma$ according to the following definition:

\begin{definition} (invertible context) A context $\Gamma$ in $\dsub$ is invertible
  if all of the following hold.
  \begin{enumerate}
  \item No variable binds to $\bot$,
  \item No variable binds to a type of the form $\{A : S .. \bot \}$ for any $S$,
  \item No variable binds to a type of the form $\{A : T .. \{A : S .. U \} \}$ for
    any $T$, $S$ and $U$, and
  \item If a variable binds to $\{A : S .. U\}$, then $S = \bot$.
  \end{enumerate}
\end{definition}

The contexts in the range of the mapping from \Cref{def:mapping} are all invertible:

\begin{lemma}
  Given an $\fsub^-$ context $\Gamma$, $\intctx{\Gamma}$ is invertible.
\end{lemma}

In invertible contexts, we can prove many useful inversion properties:

\begin{lemma}\label{lem:inv-sub-props}
  (supertypes in invertible contexts) If a context
  $\Gamma$ is invertible, then all of the following hold.
  \begin{enumerate}
  \item If $\subtypingd \top T$, then $T = \top$.
  \item If $\subtypingd{\{A: S .. U \}} T$, then $T = \top$ or $T$ has the form 
    $\{A: S' .. U'\}$.
  \item If $\subtypingd{\forall(x: S) U}T$, then $T = \top$ or $T$ has the form 
    $\forall(x : S') U'$.
  \end{enumerate}
\end{lemma}

\begin{lemma}\label{lem:inv-inv-props}
  (subtypes in invertible contexts) If a context $\Gamma$
  is invertible, then all of the following hold.
  \begin{enumerate}
  \item If $\subtypingd T \bot$, then $T = \bot$.
  \item If $\subtypingd T {\{A: S .. U \}}$, then $T = \bot$ or $T$ has the form 
    $\{A: S' .. U'\}$.
  \item If $\subtypingd T {\forall(x: S) U}$, then $T = \bot$ or $T$ is some path
    $y.A$, or $T$ has the form $\forall(x : S') U'$.
  \item If $\subtypingd {T} {x.A}$, then $T = \bot$, $T = x.A$, or $T$ is some path
    $y.A$ from which $x.A$ can be reached.
  \end{enumerate}
\end{lemma}

\begin{lemma}\label{lem:inv-props}
  (subtyping inversion) If a context $\Gamma$ is
  invertible, then the following hold.
  \begin{enumerate}
  \item If $\subtypingd{\{A : S_1 .. U_1\}}{\{A : S_2 .. U_2\}}$, then
    $\subtypingd{S_2}{S_1}$ and $\subtypingd{U_1}{U_2}$. 
  \item If $\subtypingd{\forall(x : S_1)U_1}{\forall(x : S_2)U_2}$, then
    $\subtypingd{S_2}{S_1}$ and $\subtypingd[\Gamma ; x : S_2]{U_1}{U_2}$.
  \end{enumerate}
\end{lemma}

These lemmas show that $\dsub$ is getting much closer to
$\fsub^-$ in invertible contexts (hence in the contexts in the range of $\intctx{\cdot}$)
and suggest that we are just one step away from proving
undecidability of $\dsub$. Unfortunately, there is one more problem.

\subsection{The \infref{dsub:trans} Rule}\label{ssec:attempt2}

%

Recall the conjecture that we are trying to prove:
if $\subtypingd[\intctx{\Gamma}]{\inttyp S}{\inttyp U}$,
then $\subtypingfm S U$. When we perform induction on
the premise, in the case of the \infref{dsub:trans} rule,
we have the following antecedents in the proof context:
\begin{enumerate}
\item\label[premise]{enum:s-sub-t} For some $T$, $\subtypingd[\intctx{\Gamma}]{\inttyp{S}} T$,
\item\label[premise]{enum:t-sub-u} $\subtypingd[\intctx{\Gamma}]T{\inttyp{U}}$,
\item\label[premise]{enum:ind-hyp} Inductive hypothesis: if
  $\subtypingd[\intctx{\Gamma}]{\inttyp{T_1}}{\inttyp{T_2}}$, then
  $\subtypingfm{T_1}{T_2}$.
\end{enumerate}

The problem is that $T$ is not necessarily in the range of $\inttyp{\ }$.

\paragraph{A counterexample for the \infref{dsub:trans} case: } Define $x <: T$ as
syntactic sugar for $x : \{A: \bot .. T\}$. The following is a proof that
Merlin gives us to prove that $\subtypingd[]{\forall(x <: \top) \top}{\top}$:
\begin{mathpar}
  \inferrule*[right=\infref{dsub:trans}]
  {\inferrule*[]
  {\mathcal{D}}
  {\subtypingd[]{\forall(x <: \top) \top}{\forall(x : \{A: \top .. \bot\}) x.A}} \\
  \inferrule*[right=\infref{dsub:top}]
  { }
  {\subtypingd[]{\forall(x : \{A: \top .. \bot\}) x.A}\top}
  }
  {\subtypingd[]{\forall(x <: \top) \top}{\top}}
\end{mathpar}

In the above, the subderivation 
$\mathcal{D}$ is as shown below.
\begin{mathpar}
  \inferrule*[right=\infref{dsub:all}]
  {\inferrule*[right=\infref{dsub:bnd}]
    {\text{straightforward}}
    {\subtypingd[]{\{A: \top .. \bot\}}{\{A: \bot .. \top\}}} \\
    \inferrule*[right=\infref{dsub:sel1}]
    {\text{straightforward}}
    {\subtypingd[x:\{A: \top .. \bot\}]\top{x.A}}}
  {\subtypingd[]{\forall(x <: \top) \top}{\forall(x : \{A: \top .. \bot\}) x.A}}
\end{mathpar}

In this example, the proof of $\subtypingd[]{\forall(x <: \top) \top}{\top}$ is
concluded by transitivity on $\forall(x : \{A: \top .. \bot\}) x.A$. An inspection
shows that both $\forall(x <: \top) \top$ and $\top$ are in the image of $\inttyp{\cdot}$,
but $\forall(x : \{A: \top .. \bot\}) x.A$ is not, as it would require at the very
least the lower bound in the type declaration to be $\bot$. Therefore, the target
theorem cannot be proven by induction, because the induction hypothesis can be
applied only to types in the range of $\inttyp{\cdot}$.
To resolve this issue, we need to reformulate $\dsub$ so that it does not
use the \infref{dsub:trans} rule.

\subsection{$\dsub$ Normal Form}\label{ssec:nf}

Although $\fsub$ also has a \infref{fsub:trans} rule,
it does not cause any problems for the undecidability proof of~\citet{fsub-undec}.
The reason is
that the paper begins with $\fsub$ normal form, a formulation that defines the
same calculus but does not use the
\infref{fsub:trans} rule. Therefore, it
is interesting to ask whether there is $\dsub$ normal form. We first define what we mean
by a normal form.

\begin{definition}
  A subtyping definition is in normal form if the premises of every rule are defined
  in terms of syntactic subterms of the conclusion.
\end{definition}

The subterms of the conclusion $\subtyping S U$ include subterms of both $S$ and $U$, as well as of the context $\Gamma$. Consider the following
rule in $\fsub$ normal form. 
\begin{mathpar}
  \inferrule*[right=\infref{fsubnf:tvar}]
  {X <: T \in \Gamma \\ \subtypingf T U}
  {\subtypingf X U}
\end{mathpar}

Although $T$ is not found in $X$ or $U$, notice that $T$ is a result
of context lookup of $X$ and is therefore a subterm of the context $\Gamma$.

Consider the \infref{fsub:trans} rule in $\fsub$.
\begin{mathpar}
  \inferrule*[right=\infref{fsub:trans}]
  {\subtypingf S T \\ \subtypingf T U}
  {\subtypingf S U}
\end{mathpar}

In this rule, $T$ could be arbitrary and is unrelated to the inputs. Therefore, a definition
in normal form should not contain rules like this.

We have discovered a reformulation of the $\dsub$ subtyping relation that is in normal
form. The normal form subtyping rules are shown in \Cref{fig:dsub_nf}.
The difference from the original $\dsub$ rules is that
the \infref{dsub:trans} rule is removed and replaced by the new \infref{dsub:bb}
rule.
By inspecting the rules one by one, we can see that they are indeed in normal form.
We must also check that the normal form rules define the same subtyping relation
as the original $\dsub$ subtyping rules. As a first step, we will show that
$\dsub$ normal form satisfies transitivity.

\begin{figure}
  \begin{mathpar}
    \inferrule*[right=\infref{dsub:top}]
    { }{\subtypingd{T}\top}

    \inferrule*[right=\infref{dsub:bot}]
    { }{\subtypingd \bot T}

    \inferrule*[right=\infref{dsub:refl}]
    { }{\subtypingd T T}

    \inferrule*[right=\infref{dsub:bnd}]
    {\subtypingd{S_2}{S_1} \\ \subtypingd{U_1}{U_2}}
    {\subtypingd{\{ A: S_1 .. U_1 \}}{\{A:S_2 .. U_2\}}}
    
    \inferrule*[right=\infref{dsub:all}]
    {\subtypingd{S_2}{S_1} \\ \subtypingd[\Gamma; x : S_2]{U_1}{U_2}}
    {\subtypingd{\forall(x: S_1)U_1}{\forall(x: S_2)U_2}}

    \inferrule*[right=\infref{dsub:sel1p}]
    {\subtypingd{\Gamma(x)}{\{A: S .. \top\}}}
    {\subtypingd{S}{x.A}}
    
    \inferrule*[right=\infref{dsub:sel2p}]
    {\subtypingd{\Gamma(x)}{\{A: \bot .. U\}}}
    {\subtypingd{x.A}{U}}

    \colorbox{light-gray}{
      $\inferrule*[right=\infnlabel{BB}{dsub:bb}]
      {\subtypingd{\Gamma(x)}{\{A : S .. \top\}} \\
        \subtypingd{\Gamma(x)}{\{A : \bot .. U\}} \\ \text{(for some $x$)}}
      {\subtypingd{S}{U}}$
    }
  \end{mathpar}
  
  \caption{Definition of subtyping of $\dsub$ normal form}\label{fig:dsub_nf}
\end{figure}

Proving transitivity of $\dsub$ normal form is quite tricky. First, transitivity
is interdependent with narrowing, so we will need to prove the two together.
Second, the proof of transitivity requires reasoning about types of the
following form:
$$\{A : T_1 .. \{A: T_2 .. \{A: T_3 .. \cdots \{A: T_n .. T\} \cdots \} \} \}$$
We define such types formally as follows:
\begin{definition}
A type declaration hierarchy is a type $\tthier(T, l)$ defined by another type $T$ and a list of
types $l$ inductively as follows.
\begin{align*}
    \tthier(T, l) = \begin{cases}
    T, \text{if $l$ = nil, or} \\
    \{A : T' .. \tthier(T, l')\}, \text{if $l = T' :: l'$}
    \end{cases}
\end{align*}
\end{definition}

With that definition, we can now state and prove the full transitivity and narrowing theorem:


\begin{theorem}\label{thm:dsub-trans-narr}
  For any type $T$ and two subtyping derivations in $\dsub$ normal form $\cD_1$ and
  $\cD_2$, the following hold:
  \begin{enumerate}[(1)]
  \item\label{itm:trans} (transitivity) If $\cD_1$ concludes $\subtypingd S T$
    and $\cD_2$ concludes $\subtypingd T U$, then $\subtyping S U$.
  \item\label{itm:narrow} (narrowing) If $\cD_1$ concludes $\subtypingd S T$
    and $\cD_2$ concludes $\subtypingd[\Gamma ; x : T ; \Gamma'] {S'} {U'}$, then
    $\subtypingd[\Gamma ; x : S ; \Gamma'] {S'} {U'}$.
  \item\label{itm:hdt-r} If $\cD_1$ concludes
    $\subtypingd{T'}{\tthier(\{A : S' .. T\}, l)}$ and $\cD_2$ concludes
    $\subtypingd T U$, then $\subtypingd{T'}{\tthier(\{A : S' .. U\}, l)}$.
  \item\label{itm:hdt-l} If $\cD_1$ concludes $\subtypingd S T$ and $\cD_2$
    concludes $\subtypingd{T'}{\tthier(\{A : T .. U' \}, l)}$, then
    $\subtypingd{T'}{\tthier(\{A : S .. U' \}, l)}$.
    \end{enumerate}
\end{theorem}
\begin{proof}
  The proof is done by induction on the lexicographical order of the structure of the
  triple $(T, \cD_1, \cD_2)$. That is, the inductive hypotheses of the theorem are:
  \begin{enumerate}[(a)]
  \item\label{itm:ind-t} If $T^*$ is a strict syntactic subterm of $T$, then the
    theorem holds for $T^*$ and any other two subtyping derivations $\cD'_1$ and
    $\cD'_2$.
  \item\label{itm:ind-1} If $\cD^*_1$ is a strict subderivation of $\cD_1$, then the
    theorem holds for the same type $T$, the subderivation $\cD^*_1$ and any
    subtyping derivation $\cD'_2$.
  \item\label{itm:ind-2} If $\cD^*_2$ is a strict subderivation of $\cD_2$, then the
    theorem holds for the same type $T$, the same derivation $\cD_1$ and the
    subderivation $\cD^*_2$.
  \end{enumerate}
  This form of induction is motivated by the dependencies between the four clauses of
  the theorem and can be found in other literature~\citep[Theorem 5
  (Cut)]{PFENNING200084}.  Specifically, \ref{itm:ind-t} addresses that
  transitivity~\ref{itm:trans} and narrowing~\ref{itm:narrow} are mutually dependent,
  but when transitivity uses narrowing, $T$ is replaced with a syntactic subterm
  $T^*$. Similarly, \ref{itm:ind-1} addresses that transitivity~\ref{itm:trans}
  and~\ref{itm:hdt-r} are mutually dependent, but in each dependence cycle, $\cD_1$ is
  replaced with a subderivation $\cD^*_1$. Finally, \ref{itm:ind-2} addresses that
  transitivity~\ref{itm:trans} and~\ref{itm:hdt-l} are mutually dependent, but in each
  dependence cycle, $\cD_2$ is replaced with a subderivation $\cD^*_2$.

  In proving transitivity \ref{itm:trans}, we consider the cases by which
  $\subtypingd{S}{T}$ and $\subtypingd{T}{U}$ are derived. We consider three cases in
  detail:

  \infref{dsub:all}-\infref{dsub:all} case: In this case, $S$, $T$ and $U$ are all
  dependent function types. Let $S = \forall(x : S_1)U_1$, $T = \forall(x : S_2)U_2$
  and $U = \forall(x : S_3)U_3$. The antecedents are:
  \begin{enumerate}[i.]
  \item\label{ante:s2s1} $\subtypingd{S_2}{S_1}$,
  \item\label{ante:s3s2} $\subtypingd{S_3}{S_2}$,
  \item\label{ante:u1u2} $\subtypingd[\Gamma;x : S_2]{U_1}{U_2}$, and
  \item\label{ante:u2u3} $\subtypingd[\Gamma;x : S_3]{U_2}{U_3}$.
  \end{enumerate}
  The goal is to show $\subtypingd{\forall(x : S_1)U_1}{\forall(x : S_3)U_3}$ by
  \infref{dsub:all}, which requires $\subtypingd{S_3}{S_1}$ and
  $\subtypingd[\Gamma ; x : S_3]{U_1}{U_3}$. Applying inductive
  hypothesis~\ref{itm:ind-t} to~\ref{ante:s3s2} and \ref{ante:s2s1}, we obtain
  $\subtypingd{S_3}{S_1}$ via transitivity~\ref{itm:trans}. Again applying inductive
  hypothesis~\ref{itm:ind-t} to \ref{ante:s3s2} and \ref{ante:u1u2}, we obtain
  $\subtypingd[\Gamma;x : \highlight{S_3}]{U_1}{U_2}$ via
  narrowing~\ref{itm:narrow}. Then, again by inductive hypothesis~\ref{itm:ind-t} and
  \ref{ante:u2u3}, $\subtypingd[\Gamma ; x : S_3]{U_1}{U_3}$ is concluded via
  transitivity~\ref{itm:trans} and the goal is also concluded.

  \infref{dsub:sel1p}-\infref{dsub:sel2p} case: In this case, we know $T = x.A$ for
  some $x$. The antecedents are:
  \begin{enumerate}[i.]
      \item $\subtypingd{\Gamma(x)}{\{A : S .. \top\}}$ and
      \item $\subtypingd{\Gamma(x)}{\{A : \bot .. U\}}$.
  \end{enumerate}
  By the \infref{dsub:bb} rule, we can show the conclusion $\subtypingd S U$.  That
  is, the \infref{dsub:bb} rule is a restricted form of transitivity for the case when
  the middle type is a path type $x.A$.

  \infref{dsub:sel2p}-\emph{any} case: When $\subtypingd S T$ is derived by
  \infref{dsub:sel2p}, we know $S = y.A$ for some $y$.  The antecedents are:
  \begin{enumerate}[i.]
  \item $\subtypingd{\Gamma(y)}{\{A : \bot .. T\}}$, and
  \item $\subtypingd T U$.
  \end{enumerate}
  The intention is to show that
  $\subtypingd{\Gamma(y)}{\{A : \bot .. \highlight{U}\}}$ holds and hence
  conclude $\subtypingd{y.A}{U}$ by \infref{dsub:sel2p}. To derive this conclusion, we
  need to apply the induction hypothesis~\ref{itm:ind-1} with
  $\subtypingd{\Gamma(y)}{\{A : \bot .. T\}}$ as the subderivation $\cD^*_1$. The
  induction hypothesis~\ref{itm:ind-1} provides the necessary
  $\subtypingd{\Gamma(y)}{\{A : \bot .. U\}}$ via clause~\ref{itm:hdt-r}, and hence
  $\subtypingd{y.A}{U}$. The \infref{dsub:bb}-\emph{any} case can be proved in the
  same way. The \emph{any}-\infref{dsub:sel1p} and \emph{any}-\infref{dsub:bb} cases
  can be proved in a symmetric way, by invoking inductive hypothesis~\ref{itm:ind-2}
  instead of inductive hypothesis~\ref{itm:ind-1} in the corresponding places.
  
  Narrowing \ref{itm:narrow} is proved by case analysis on the derivation of
  $\subtypingd[\Gamma ; x : T ; \Gamma']{S'}{U'}$. Several cases require transitivity,
  which is obtained by applying the induction hypothesis~\ref{itm:ind-2}.

  Clause~\ref{itm:hdt-r} of the theorem is proved by case analysis on $\cD_1$, the
  derivation of $\subtypingd{T'}{\tthier(\{A:S'..T\},l)}$, and then by an inner
  induction on the list $l$.  We discuss two interesting cases.

  \infref{dsub:bnd}-nil case: $\tthier(\{A : S' .. T\}, nil) = \{A : S' .. T\}$ and
  $\subtypingd{T'}{\tthier(\{A : S' .. T\}, nil)}$ is constructed by
  \infref{dsub:bnd}. From the \infref{dsub:bnd} rule, we know that
  $T' = \{A : S_0 .. U_0\}$ and have the following antecedents:
  \begin{enumerate}[i.]
  \item $\subtypingd{S'}{S_0}$, and
  \item\label{ante:u0t} $\subtypingd{U_0}{T}$, and
  \item\label{ante:tu} $\subtypingd{T}{U}$.
  \end{enumerate}

  We wish to apply transitivity~\ref{itm:trans} to antecedents~\ref{ante:u0t} and
  \ref{ante:tu} to obtain $\subtypingd{U_0}{U}$. We can do this by invoking the
  induction hypothesis~\ref{itm:ind-1} with the antecedent~\ref{ante:u0t}
  $\subtypingd{U_0}{T}$ as $\cD^*_1$. After applying transitivity, we can apply
  \infref{dsub:bnd} to $\subtypingd{S'}{S_0}$ and $\subtypingd{U_0}{U}$ to obtain
  $\subtypingd{\{A : S_0 .. U_0\}}{\{A : S' .. U\}}$ as required.  This case shows the
  mutual dependence between clause~\ref{itm:hdt-r} and transitivity~\ref{itm:trans}.

  \infref{dsub:sel2p}-\emph{any} case: In this case, we know that $T' = z.A$ for some
  $z$ and have the following antecedents:
  \begin{enumerate}[i.]
  \item $\subtypingd{\Gamma(z)}{\{A : \bot .. \tthier(\{A : S' .. T\}, l)\}}$, and
  \item $\subtypingd T U$.
  \end{enumerate}

  We apply the induction hypothesis~\ref{itm:ind-1} with the subderivation
  $\subtypingd{\Gamma(z)}{\{A : \bot .. \tthier(\{A : S' .. T\}, l)\}}$ as $\cD^*_1$.
  Notice that $\{A : \bot .. \tthier(\{A : S' .. T\}, l)\}$ can be rewritten as
  $\tthier(\{A : S' .. T\}, (\bot :: l))$, so the induction hypothesis
  of~\ref{itm:hdt-r} applies to yield
  $\subtypingd{\Gamma(z)}{\tthier(\{A : S' .. \highlight{U}\}, (\bot :: l))}$, which
  can be rewritten as
  $\subtypingd{\Gamma(z)}{\{A: \bot .. \tthier(\{A : S' .. U\}, l)\}}$.  Finally, by
  \infref{dsub:sel2p}, $\subtypingd{z.A}{\tthier(\{A : S' .. U\}, l)}$ as required.
  Since the list $\bot :: l$ is longer than $l$, this case shows why
  clause~\ref{itm:hdt-r} needs to be defined on type declaration hierarchies of
  non-empty lists.

  Clause~\ref{itm:hdt-l} of the theorem is dual to clause~\ref{itm:hdt-r} and is
  proven in a symmetric way. Instead of the inductive hypothesis~\ref{itm:ind-1},
  clause~\ref{itm:hdt-l} uses the inductive hypothesis~\ref{itm:ind-2}.
\end{proof}

Once transitivity is proved, we can show that the two definitions of $\dsub$ subtyping are
equivalent.

\begin{theorem}\label{thm:dsub-nf-equiv}
  Subtyping in $\dsub$ normal form is equivalent to the original $\dsub$.
\end{theorem}
\begin{proof}
  The if direction is immediate. In the only if direction, the \infref{dsub:trans}
  case can be discharged by transitivity of $\dsub$ normal form.
\end{proof}

Now that we have $\dsub$ normal form, we can finally show that $\dsub$ subtyping
is indeed undecidable.

\begin{theorem}
    \label{thm:dsub2fsubm}
  If $\subtypingd[\intctx{\Gamma}]{\inttyp{S}}{\inttyp{U}}$, then $\subtypingfm S U$. 
\end{theorem}
\begin{proof}
  The proof is by induction on the subtyping derivation in $\dsub$ normal form,
  which no longer has the problem with the \infref{dsub:trans} rule discussed
  in \Cref{ssec:attempt2}. 
  Most of the cases are proved by straightforward application of the induction hypothesis. 
  The \infref{dsub:sel1p} and \infref{dsub:bb} cases require the following argument.
  In both cases, we have the antecedent:
  \begin{mathpar}
    \text{for some $x$, }\subtypingd[\intctx{\Gamma}]{\intctx{\Gamma}(x)}{\{A : \inttyp{S}
      .. \top\}}
  \end{mathpar}

  By inspecting $\intctx{\cdot}$, we know that $\intctx{\Gamma}(x)$ must be $\{A : \bot
  .. T\}$ for some $T$, and therefore the antecedent becomes
  \begin{mathpar}
    \subtypingd[\intctx{\Gamma}]{\{A : \bot .. T\}}{\{A : \inttyp{S} .. \top\}}
  \end{mathpar}
  Recall that $\intctx{\Gamma}$ is invertible. By \Cref{lem:inv-props}, we know
  \begin{mathpar}
    \subtypingd[\intctx{\Gamma}]{\inttyp{S}}\bot
  \end{mathpar}
  Furthermore, by \Cref{lem:inv-inv-props}, we know $\inttyp{S} = \bot$. By inspecting
  $\inttyp{\cdot}$, we see that $\bot$ is not in the image, and therefore both
  \infref{dsub:sel1p} and \infref{dsub:bb} cases are discharged by contradiction.
\end{proof}

\begin{theorem}
    \label{thm:dsubundec}
  Subtyping in $\dsub$ is undecidable.
\end{theorem}
\begin{proof}
    The proof is by reduction from $\fsubm$ using the mapping from
    \Cref{def:mapping} but without the function case. For the if direction,
    \Cref{thm:fsub2dsub} applies since the $\fsubm$ subtyping rules are a subset of
    the $\fsub$ subtyping rules. The only if direction is proved by the previous theorem.
\end{proof}

As we have seen, the only change in the normal form rules of $\dsub$ subtyping
is that the \infref{dsub:trans} rule is removed and replaced with the \infref{dsub:bb}
rule.
In
other words, the only thing that transitivity really contributes to $\dsub$ is the phenomenon of
bad bounds. Conversely, if we exclude bad bounds from $\dsub$, then it no longer
has transitivity of subtyping.

The undecidability proof relies only on the common features of $\fsubm$ and $\dsub$,
and in particular, it does not depend on the \infref{dsub:bb} rule. If we remove
this rule from $\dsub$, subtyping in the resulting variant is still undecidable.

\begin{theorem}
  Subtyping in $\dsub$ normal form without the \infref{dsub:bb} rule is undecidable.
\end{theorem}
\begin{proof}
    The proof is the same as \Cref{thm:dsubundec}, but without the \infref{dsub:bb} case.
\end{proof}

\subsection{Undecidability of Typing}

In most calculi, undecidability of typing usually follows by some simple
reduction from undecidability of subtyping in the same calculus. For example,
for $\dsub$, we might map the subtyping problem
$\subtypingd S U$
to the typing problem:
\begin{mathpar}
  \typingd{\{A = S\}}{\{A : \bot .. U\}}
\end{mathpar}
and conjecture that the two problems are equivalent. In $\dsub$, however, we have to
be careful because of the possibility of bad bounds. Indeed, it turns out that the two problems are \emph{not} equivalent. As
a counterexample, note that if 
$\Gamma(w) = \{A : \{A : S .. S\} .. \{A : \bot .. U\} \}$, then the typing problem is
true (since 
$\typingd{\{A = S\}}{\{A : S .. S\}}$ and 
$\subtypingd {\{A : S .. S\}}{\{A : \bot .. U\}}$)
even if $S$ and $U$ are chosen so that the subtyping problem $\subtypingd S U$ is false.

In general, the approach to proving undecidability of typing using undecidability of subtyping depends on inversion properties, which do not always hold in
$\dsub$ due to bad bounds, so this approach does not work for $\dsub$.
Nevertheless, $\dsub$ typing still turns out to be undecidable, but to prove it, we must reduce not from
$\dsub$ subtyping, but from $\fsubm$ subtyping, which does obey inversion properties.
\begin{theorem}
  For all $\Gamma$, $S$ and $U$ in $\fsub^-$, 
  \begin{mathpar}
    \text{if }\typingd[\intctx{\Gamma}]{\{A = \inttyp{S}\}}{\{A : \bot .. \inttyp{U}\}}\text{, then }
    \subtypingfm{S}{U}\text{.}
  \end{mathpar}
\end{theorem}
\begin{proof}
    The only typing rules that apply to $\{A = \inttyp{S}\}$ are
\infref{dsub:typ-i} and $\infref{dsub:sub}$. Therefore, the premise
implies that
$\subtypingd[\intctx{\Gamma}]{\{A : \inttyp{S} .. \inttyp{S}\}}{\{A : \bot
  .. \inttyp{U}\}}$. Since $\intctx{\Gamma}$ is invertible, \Cref{lem:inv-props} implies
  $\subtypingd[\intctx{\Gamma}]{\inttyp{S}}{\inttyp{U}}$
  and \Cref{thm:dsub2fsubm} implies
    $\subtypingfm{S}{U}$.
\end{proof}

\begin{theorem}
  $\dsub$ typing is undecidable.
\end{theorem}

\begin{proof}
    By reduction from $\fsubm$ subtyping, mapping the $\fsubm$ subtyping problem $\subtypingfm{S}{U}$ to the $\dsub$ typing problem
    $\typingd[\intctx{\Gamma}]{\{A = \inttyp{S}\}}{\{A : \bot .. \inttyp{U}\}}$.
    The if direction is immediate and the only if direction is proved by the previous theorem.
\end{proof}

\section{Kernel $\dsub$}

\subsection{Motivation and Definition}\label{ssec:kdsub-def}

In the previous section, we showed that both typing and subtyping in $\dsub$ are
undecidable. A natural question to ask is \emph{what fragments of $\dsub$ are decidable?}
In this section, we consider one such fragment.

We base our adjustments to $\dsub$ on its normal form. The first adjustment is
inspired by $\fsub$, which becomes decidable if its \infref{fsub:all} rule is
restricted to a \emph{kernel} rule that requires the parameter types of both universal
types to be identical~\cite{Cardelli:1985:UTD:6041.6042}. We apply the same
restriction to the $\dsub$ \infref{dsub:all} rule.

The second adjustment is to remove the \infref{dsub:bb} rule. There are several
reasons for that:
\begin{enumerate}
\item Bad bounds are consequences of unintended interactions between the
  \infref{dsub:sel1p}, \infref{dsub:sel2p} and \infref{dsub:trans} rules.
\item \citet{abel-algorithmic} observed that the implementation of subtyping in the Scala compiler
    violates transitivity in some cases, and these cases correspond exactly
    to the \infref{dsub:bb} rule. That is, the Scala compiler does not implement
    this rule.
\item We conjecture that a calculus with bad bounds will be undecidable.
\end{enumerate}

The calculus after these two changes is shown in \Cref{fig:dsub_kernel}.
We will see that this calculus is decidable, so
we call it \emph{kernel $\dsub$}, following
the convention in \cite{tapl}.

\begin{figure}
  \begin{mathpar}
    \inferrule*[right=\infnlabel{K-Top}{dsub:k:top}]
    { }{\subtypingdk{T}\top}

    \inferrule*[right=\infnlabel{K-Bot}{dsub:k:bot}]
    { }{\subtypingdk \bot T}

    \inferrule*[right=\infnlabel{K-VRefl}{dsub:k:vrefl}]
    { }{\subtypingdk{x.A}{x.A}}

    \inferrule*[right=\infnlabel{K-Bnd}{dsub:k:bnd}]
    {\subtypingdk{S_2}{S_1} \\ \subtypingdk{U_1}{U_2}}
    {\subtypingdk{\{ A: S_1 .. U_1 \}}{\{A:S_2 .. U_2\}}}
    
    \inferrule*[right=\infnlabel{K-All}{dsub:k:all}]
    {\subtypingdk[\Gamma; x : S]{U_1}{U_2}}
    {\subtypingdk{\forall(x: S)U_1}{\forall(x: S)U_2}}

    \inferrule*[right=\infnlabel{K-Sel1}{dsub:k:sel1}]
    {\subtypingdk{\Gamma(x)}{\{A: S .. \top\}}}
    {\subtypingdk{S}{x.A}}
    
    \inferrule*[right=\infnlabel{K-Sel2}{dsub:k:sel2}]
    {\subtypingdk{\Gamma(x)}{\{A: \bot .. U\}}}
    {\subtypingdk{x.A}{U}}
  \end{mathpar}
  
  \caption{Definition of kernel $\dsub$}\label{fig:dsub_kernel}
\end{figure}

%
We can show that kernel $\dsub$ is sound with respect to the original (full) $\dsub$:

\begin{theorem}
  If $\subtypingdk S U$, then $\subtypingd S U$. 
\end{theorem}

If kernel $\dsub$ is decidable, it cannot also be complete for full $\dsub$. For
example, it does not admit the following subtyping judgment that is admitted by full
$\dsub$.
\begin{mathpar}
  \subtypingd[x : \{A : \top .. \top\}]{\forall(y : x.A)\top}{\forall(y : \top)\top}
\end{mathpar}

Kernel $\dsub$ rejects it because $x.A$ and $\top$ are not syntactically identical.

Moreover, kernel $\dsub$ rejects conclusions that can only be drawn from bad
bounds, such as:
\begin{mathpar}
  \subtypingd[x : \{A : \top .. \bot\}]\top \bot
\end{mathpar}
This judgment can only be achieved by invoking \infref{dsub:trans} or
\infref{dsub:bb}, but both of these rules are absent from kernel $\dsub$.

\subsection{Step subtyping}\label{ssec:step}

\citet{abel-algorithmic} defined step subtyping, a partial algorithm for deciding a
fragment of $\dsub$ subtyping based on ideas developed for subtyping in $\fsub$
by~\citet{tapl}. We
briefly review the step subtyping algorithm here.  In the next section, we will
observe that the fragment of $\dsub$ subtyping decided by the algorithm turns out to
be exactly the kernel $\dsub$ that we defined in the previous section.  We made some
adjustments to the presentation to set up a framework, so the definition is not
identical to Nieto's, but the adjustments are minor and have no impact on
expressiveness.

\begin{figure}
  \begin{mathpar}
    \inferrule*[right=\infnlabel{S-Top}{dsub:step:top}]
    { }
    {\subtypingda T \top}
    
    \inferrule*[right=\infnlabel{S-Bot}{dsub:step:bot}]
    { }
    {\subtypingda \bot T}

    \inferrule*[right=\infnlabel{S-VRefl}{dsub:step:vrefl}]
    { }
    {\subtypingda{x.A}{x.A}}

    \inferrule*[right=\infnlabel{S-Bnd}{dsub:step:bnd}]
    {\subtypingda{S'} S \\ \subtypingda U {U'}}
    {\subtypingda{\{A : S .. U\}}{\{A : S' .. U'\}}}

    \inferrule*[right=\infnlabel{S-All}{dsub:step:all}]
    {\\ \subtypingda[\Gamma ; x : S] U {U'}}
    {\subtypingda{\forall(x : S) U}{\forall(x : S) U'}}

    \inferrule*[right=\infnlabel{S-Sel1}{dsub:step:sel1}]
    {\dced{x.A} S \\ \subtypingda T S}
    {\subtypingda{T}{x.A}}

    \inferrule*[right=\infnlabel{S-Sel2}{dsub:step:sel2}]
    {\uced{x.A} U \\ \subtypingda U T}
    {\subtypingda{x.A}{T}}
  \end{mathpar}

  \caption{Definition of step subtyping
    operation~\citep{abel-algorithmic}}\label{fig:dsub-step-sub} 
\end{figure}

\begin{figure}
  \exposure
  \begin{mathpar}
    \inferrule*[right=\infnlabel{Exp-Stop}{dsub:expo:stop}]
    {T\text{ is not a path}}
    {\expod T T}

    \inferrule*[right=\infnlabel{Exp-Top}{dsub:expo:top}*]
    { }
    {\expod T \top}

    \inferrule*[right=\infnlabel{Exp-Bot}{dsub:expo:bot}]
    {\expod[\Gamma_1] T \bot}
    {\expod[\Gamma_1 ; x : T ; \Gamma_2]{x.A} \bot}

    \inferrule*[right=\infnlabel{Exp-Bnd}{dsub:expo:bnd}]
    {\expod[\Gamma_1] T {\{A : S .. U\}} \\ \expod[\Gamma_1]{U}{U'}}
    {\expod[\Gamma_1 ; x : T ; \Gamma_2]{x.A}{U'}}    
  \end{mathpar}

  \upcast / \downcast
  \begin{mathpar}
    \inferrule*[right=\infnlabel{Uc-Top}{dsub:uc:top}*]
    { }
    {\uced{x.A}\top}

    \inferrule*[right=\infnlabel{Uc-Bot}{dsub:uc:bot}]
    {\expod[\Gamma_1] T \bot}
    {\uced[\Gamma_1 ; x : T ; \Gamma_2]{x.A}\bot}

    \inferrule*[right=\infnlabel{Uc-Bnd}{dsub:uc:bnd}]
    {\expod[\Gamma_1] T {\{A : S .. U\}}}
    {\uced[\Gamma_1 ; x : T ; \Gamma_2]{x.A}{U}}

    \inferrule*[right=\infnlabel{Dc-bot}{dsub:dc:bot}*]
    { }
    {\dced{x.A}\bot}
    
    \inferrule*[right=\infnlabel{Dc-Top}{dsub:dc:top}]
    {\expod[\Gamma_1] T \bot}
    {\dced[\Gamma_1 ; x : T ; \Gamma_2]{x.A}\top}

    \inferrule*[right=\infnlabel{Dc-Bnd}{dsub:dc:bnd}]
    {\expod[\Gamma_1] T {\{A : S .. U\}}}
    {\dced[\Gamma_1 ; x : T ; \Gamma_2]{x.A}{S}}
  \end{mathpar}
  
  \caption{Definitions of \textbf{Exposure} and \textbf{Upcast / Downcast}
    operations~\citep{abel-algorithmic}}\label{fig:dsub-exposure} 
\end{figure}

\begin{definition}
  Step subtyping is defined using the inference rules in \Cref{fig:dsub-step-sub}.
  The algorithm searches for a derivation using these rules, backtracking if
  necessary. Backtracking eventually terminates by \Cref{thm:step-term}.
\end{definition}

The definitions of kernel $\dsub$ and step subtyping look similar.
The differences are the cases related to path types.
For these types, step subtyping uses three additional operations,
\exposure ($\Uparrow$), \upcast ($\nearrow$), and \downcast ($\searrow$).
The purpose of \upcast (\downcast) is, given a path type $x.A$,
to look up $x$ in the typing context to a type member declaration $\{A:S..U\}$
and read off the upper bound $U$ (lower bound $S$, respectively). A complication,
however, is that the typing context could assign to $x$ another path
type. Therefore, \upcast and \downcast use \exposure, whose purpose is to convert
a type that could be a path type to a supertype that is guaranteed to not be
a path type. \exposure
maps every non-path type to itself, and it maps a path
type $x.A$ to its supertype $U$ in a similar way as \upcast. However, $U$ could
itself be a path type, so, unlike \upcast, \exposure calls itself recursively on $U$.
This guarantees that the type returned from \exposure is never a path type.

The definitions of these operations are shown in \Cref{fig:dsub-exposure}.  The
\infref{dsub:expo:top}, \infref{dsub:uc:top} and \infref{dsub:dc:bot} rules are
defined to make the operations total functions. We mark them with asterisks to
indicate that they apply only when no other rules do, and therefore each of the three
operations has exactly one rule to apply for any given input.

\upcast and \downcast are shallow wrappers over \exposure. Notice that \upcast and
\downcast are not even recursive. When handling a path type $x.A$, they use \exposure
to find a non-path supertype of $\Gamma(x)$ and simply return bounds in the right
directions. It is possible that \upcast and \downcast return other path types.

Notice that in the \infref{dsub:expo:bot} and \infref{dsub:expo:bnd} rules, the recursive calls
continue with $\Gamma_1$, the context preceding $x$. This ensures termination of \exposure. As long as the original
context $\Gamma_1, x:T, \Gamma_2$ is well-formed, $T$ is closed in the truncated
context $\Gamma_1$.

%
%
%
%
Nieto showed that step subtyping is a sound and terminating algorithm.

\begin{theorem}\citep{abel-algorithmic}
  Step subtyping as an algorithm is sound w.r.t. full $\dsub$.
  \begin{mathpar}
    \text{If }\subtypingda S U\text{, then }\subtypingd S U
  \end{mathpar}
\end{theorem}

\begin{theorem}\citep{abel-algorithmic}\label{thm:step-term}
  Step subtyping as an algorithm terminates.
\end{theorem}

%
%




\subsection{Soundness and Completeness of Step Subtyping}\label{ssec:sck}
In this section, we will show that the subset of $\dsub$ subtyping relationships
that step subtyping discovers turns out to be exactly the relation
defined by the declarative kernel $\dsub$ rules. We begin by proving some
basic properties of kernel $\dsub$.

Although the kernel $\dsub$ subtyping reflexivity rule \infref{dsub:k:vrefl} applies only to path types, subtyping is actually reflexive for all types:
\begin{lemma} Kernel $\dsub$ subtyping is reflexive.
  \begin{align*}
    \subtypingdk T T
  \end{align*}
\end{lemma}

Since kernel $\dsub$ does not have the \infref{dsub:bb} or \infref{dsub:trans} rules,
transitivity no longer holds in general, but it does hold on $\top$
and $\bot$:
\begin{lemma}\label{lem:kernel:trans:top}
  If $\subtypingdk \top U$, then $\subtypingdk S U$. 
\end{lemma}



\begin{lemma}\label{lem:kernel:trans:bot}
  If $\subtypingdk S \bot$, then $\subtypingdk S U$. 
\end{lemma}

Comparing step subtyping with kernel $\dsub$, we will show soundness of step subtyping first and completeness second. In step
subtyping, the operations are separated into two layers. The first is the subtyping
algorithm itself and the second is \exposure, which handles path types. The proof needs
to go from the reverse direction by connecting \exposure with kernel $\dsub$ first.

\begin{lemma}\label{lem:expo-dsubk}
  If $\expod S T$ and $\subtypingdk T U$, then $\subtypingdk S U$. 
\end{lemma}
\begin{proof}
    By induction on the derivation of \exposure.
\end{proof}


We can then show that step subtyping is sound.

\begin{theorem}\label{thm:kernel:sound} (soundness of step subtyping w.r.t. kernel $\dsub$) 
  If $\subtypingda S U$, then $\subtypingdk S U$. 
\end{theorem}
\begin{proof}
  By induction on step subtyping. From the rules, we can see that kernel
  $\dsub$ and step subtyping are almost identical, except for the
  \infref{dsub:step:sel1} and \infref{dsub:step:sel2} cases. These cases can be
  discharged by expanding \upcast and \downcast and then applying
  \Cref{lem:expo-dsubk}.
\end{proof}

Now we proceed to the opposite direction, proving completeness of step subtyping.

%
%
\begin{theorem}\label{thm:completeness-kernel} (completeness of step subtyping
  w.r.t. kernel $\dsub$)
 
  If $\subtypingdk S U$, then $\subtypingda S U$. 
\end{theorem}
\begin{proof}
    The proof requires an intricate strengthening of the induction hypothesis:
  if $\subtypingdk S U$ and this derivation contains $n$ steps, then $\subtypingda S U$, and if $U$ is of the form
  ${\{A : T_1 .. T_2\}}$, then $\expod S {S'}$ for some $S'$,
  and either \begin{enumerate}
      \item $S' = \bot$ or 
      \item $S' = \{A:T'_1..T'_2\}$ for some $T'_1$ and $T'_2$
          such that
          \begin{enumerate}
              \item $\subtypingda {T_1} {T_1'}$ and 
              \item $\subtypingdk{T_2'}{T_2}$, and the number
      of steps in the derivation of $\subtypingdk{T_2'}{T_2}$ is less than or equal
      to $n$.
  \end{enumerate}
  \end{enumerate}

  The proof is by strong induction on $n$.

  To prove $\subtypingda S U$, the non-trivial cases are \infref{dsub:k:sel1} and
  \infref{dsub:k:sel2} cases; we discuss the latter.  The antecedent is
  $\subtypingdk{\Gamma(x)}{\{A : \bot .. U\}}$.  This case requires the strengthened
  induction hypothesis, since the original would only imply that
  $\subtypingda{\Gamma(x)}{\{A : \bot .. U\}}$, which is insufficient to establish
  $\subtypingda{x.A}{U}$. To establish this conclusion, we wish to apply the
  \infref{dsub:step:sel2} rule. The strengthened induction hypothesis is designed
  specifically to provide the necessary premises of this rule.

%
  It remains to prove the strengthened induction hypothesis.
  The type $U$ can have the specified form $\{A : T_1 .. T_2\}$
  in the conclusions of three rules:
 \infref{dsub:k:bot},
 \infref{dsub:k:bnd} and \infref{dsub:k:sel2}. Only the \infref{dsub:k:sel2} case is interesting.
  The conclusion of this rule forces $S = y.A$ for some $y$,
  and the antecedent is
  $\subtypingdk{\Gamma(y)}{\{A : \bot .. \{A : T_1 .. T_2\}\}}$. Applying the
  induction hypothesis to this antecedent leads to two cases:
  \begin{enumerate}
  \item When $\expod{\Gamma(y)}{\bot}$, the goal $\expod{y.A}{\bot}$ follows by
    \infref{dsub:expo:bot}. 
  \item Otherwise, for some $T_1'$ and $T_2'$, we obtain additional antecedents:
    \begin{enumerate}
    \item $\expod{\Gamma(y)}{\{A : T_1' .. T_2' \}}$,
    \item $\subtypingda{\bot}{T_1'}$, and
    \item $\subtypingdk{T_2'}{\{A : T_1 .. T_2\}}$ by a derivation with strictly fewer that $n$ steps.
    \end{enumerate}
    The intention is to apply the \infref{dsub:expo:bnd} rule, but this rule requires
    an \exposure on $T_2'$ as well. This can be achieved by applying the inductive
    hypothesis to the third antecedent again. This yields 
    $\expod{T_2'}{T_2''}$ for some $T_2''$ and this case is concluded, so we can apply
    \infref{dsub:expo:bnd} to obtain $\expod{y.A}{T_2''}$, where $T_2''$ satisfies
    the properties that the strengthened induction hypothesis requires of $S'$.
  \end{enumerate}
\end{proof}

Hence, we have shown that the subrelation of $\dsub$ subtyping induced by the step subtyping algorithm is exactly the kernel $\dsub$ subtyping relation.

\section{Strong Kernel $\dsub$}

\subsection{Motivation and Definition}\label{ssec:skdsub-motiv}

In the previous section, we defined a decidable fragment of $\dsub$, kernel
$\dsub$. Notwithstanding its decidability, it comes with obvious disadvantages. One
example is the judgment we presented in \Cref{ssec:kdsub-def}:
\begin{mathpar}
  \subtypingd[x : \{A : \top .. \top\}]{\forall(y : x.A)\top}{\forall(y : \top)\top}
\end{mathpar}

This judgment is admitted in full $\dsub$ but not kernel $\dsub$.  The latter rejects
this judgment because it requires the parameter types to be syntactically
identical. However, we can see that here $x.A$ and $\top$ are in a special situation:
$x.A$ is defined with $\top$ as both its lower and upper bounds, which makes $x.A$ an
\emph{alias} for $\top$. In Scala, we would like to be able to use aliased types
interchangeably. The kernel requirement of syntactically identical parameter types
significantly restricts the usefulness of type aliases. Hence, the
aim of this section is to (at least) lift this restriction while maintaining decidability.

The inspiration for the new calculus comes from writing out the typing context \emph{twice} in a subtyping derivation. For example,
the \infref{dsub:all} rule is:
\begin{mathpar}
  \inferrule*[right=\infref{dsub:all}]
  {\subtypingd{S'}{S} \\ \subtypingd[\Gamma; x : S']{U_x}{U'_x}}
  {\subtypingd{\forall(x : S)U}{\forall(x : S')U'}}
\end{mathpar}

Let us write the contexts twice for this rule:
\begin{mathpar}
  \inferrule*[right=\infnlabel{All-TwoContexts}{dsub:all-2g}]
  {\subtypingdt{S'}{S} \\ \subtypingdt[\Gamma; x : S']{U_x}{U'_x}[\Gamma; x : S']}
  {\subtypingdt{\forall(x : S)U}{\forall(x : S')U'}}
\end{mathpar}

Now do the same for the kernel version too:
\begin{mathpar}
  \inferrule*[right=\infnlabel{K-All-TwoContexts}{dsub:k:all-2g}]
  {\subtypingdt[\Gamma; x : S]{U_x}{U'_x}[\Gamma; x : S]}
  {\subtypingdt{\forall(x : S)U}{\forall(x : S)U'}}
\end{mathpar}

So far, both copies of the context have been the same, so the second copy is redundant.
However, comparing these two rules for a moment, we
start to see some potential for improvement. In the premise comparing $U_x <: U'_x$, the
only difference are the primes on $S$ in the typing contexts: the first rule uses $S'$
on both sides, while the second rule uses $S$ on both sides. Since $U_x$ comes from a
universal type where $x$ has type $S$, and $U'_x$ from one where $x$ has type $S'$,
what if we took the middle ground between the two rules, and added $S$ to the left
context and $S'$ to the right context?
\begin{mathpar}
  \inferrule*[right=All-AsymmetricContexts]
  {\subtypingdt{S'}{S} \\ \subtypingdt[\Gamma; x : S]{U_x}{U'_x}[\Gamma; x : S']}
  {\subtypingdt{\forall(x : S)U}{\forall(x : S')U'}}
\end{mathpar}

The new rule enables the contexts to be different, so it justifies
maintaining both contexts. But how will a calculus with this hybrid rule
behave? Will it be strictly in between the decidable kernel $\dsub$ and
the undecidable full $\dsub$ in expressiveness? Will it be decidable? We will show that the
answer to both questions is yes. 
The new hybrid rule 
allows comparison of function types with \emph{different} parameter types, and the return types are
compared in two different contexts. In particular, it admits the example judgement with the aliased
parameter types with which we began this section.

We call
this new calculus \emph{strong kernel $\dsub$}, and define it fully in \Cref{fig:dsub_skernel}.
The \infref{dsub:sk:all} rule is the only rule that enables the two contexts to diverge. All of the other rules simply copy both contexts unchanged to the premises.

\begin{figure}
  \begin{mathpar}
    \inferrule*[right=\infnlabel{Sk-Top}{dsub:sk:top}]
    { }{\subtypingdsk{T}\top}

    \inferrule*[right=\infnlabel{Sk-Bot}{dsub:sk:bot}]
    { }{\subtypingdsk \bot T}

    \inferrule*[right=\infnlabel{Sk-VRefl}{dsub:sk:vrefl}]
    { }{\subtypingdsk{x.A}{x.A}}

    \inferrule*[right=\infnlabel{Sk-Bnd}{dsub:sk:bnd}]
    {\subtypingdskr{S_1}{S_2} \\ \subtypingdsk{U_1}{U_2}}
    {\subtypingdsk{\{ A: S_1 .. U_1 \}}{\{A:S_2 .. U_2\}}}
    
    \inferrule*[right=\infnlabel{Sk-All}{dsub:sk:all}]
    {\subtypingdskr{S_1}{S_2} \\ \subtypingdsk[\Gamma_1; x : S_1]{U_1}{U_2}[\Gamma_2 ; x : S_2]}
    {\subtypingdsk{\forall(x: S_1)U_1}{\forall(x: S_2)U_2}}

    \inferrule*[right=\infnlabel{Sk-Sel1}{dsub:sk:sel1}]
    {\subtypingdskr{\{A: S .. \top\}}{\Gamma_2(x)}}
    {\subtypingdsk{S}{x.A}}
    
    \inferrule*[right=\infnlabel{Sk-Sel2}{dsub:sk:sel2}]
    {\subtypingdsk{\Gamma_1(x)}{\{A: \bot .. U\}}}
    {\subtypingdsk{x.A}{U}}
  \end{mathpar}
  
  \caption{Definition of strong kernel $\dsub$}\label{fig:dsub_skernel}
\end{figure}

\subsection{Properties of Strong Kernel $\dsub$}

In this section, we will prove that the subtyping relation defined by strong kernel $\dsub$
is in between kernel $\dsub$ and full $\dsub$ in expressiveness. As a first step, we need
to prove reflexivity.

\begin{lemma} Strong kernel $\dsub$ is reflexive.
  \begin{align*}
    \subtypingdsk T T
  \end{align*}
\end{lemma}

\begin{proof}
  By induction on $T$. 
\end{proof}

In the next two theorems, we wish to show that strong kernel $\dsub$ is in between
kernel $\dsub$ and full $\dsub$ in terms of expressiveness: 

\begin{theorem}\label{thm:kdsub2skdsub}
  If $\subtypingdk S U$ then $\subtypingdsk[\Gamma]S U[\Gamma]$.
\end{theorem}
\begin{proof}
  By induction on the derivation. The \infref{dsub:k:all} case requires reflexivity of
  strong kernel $\dsub$.
\end{proof}


\begin{theorem}\label{thm:skdsub2dsub}
  If $\subtypingdsk[\Gamma]S U[\Gamma]$ then $\subtypingd S U$. 
\end{theorem}

Before we can prove this theorem, we need to define a new concept, a relationship
between the two typing contexts.

\begin{definition}\label{def:opesub}
  The order preserving sub-environment relation between two contexts, or $\opesub$, is
  defined in \Cref{fig:opesub_def}.
\end{definition}

\begin{figure}
  \begin{mathpar}
    \inferrule*[right=\infnlabel{Ope-Nil}{dsub:ope:nil}]
    { }
    {\opesubd{\bigcdot}[\bigcdot]}

    \inferrule*[right=\infnlabel{Ope-Drop}{dsub:ope:drop}]
    {\opesubd{\Gamma}}
    {\opesubd{\Gamma ; x : T}[\Gamma']}

    \inferrule*[right=\infnlabel{Ope-Keep}{dsub:ope:keep}]
    {\opesubd{\Gamma} \\ \subtypingd S U}
    {\opesubd{\Gamma ; x : S}[\Gamma' ; x : U]}
  \end{mathpar}

  \caption{Definition of $\opesub$}\label{fig:opesub_def}
\end{figure}

Intuitively, If $\opesubd{\Gamma}$, then $\Gamma$ is a more ``informative'' context
than $\Gamma'$. $\opesub$ is a combination of the narrowing and weakening properties. 
The following properties of $\opesub$ confirm this intuition.

\begin{lemma} $\opesub$ is reflexive.
  \begin{align*}
    \opesubd{\Gamma}[\Gamma]
  \end{align*}
\end{lemma}

\begin{lemma} $\opesub$ is transitive.
  \begin{align*}
    \text{If }\opesubd{\Gamma_1}[\Gamma_2]\text{ and
    }\opesubd{\Gamma_2}[\Gamma_3]\text{, then }\opesubd{\Gamma_1}[\Gamma_3].
  \end{align*}
\end{lemma}

\begin{theorem} (respectfulness) Full $\dsub$ subtyping is preserved by $\opesub$.

  If $\opesubd{\Gamma}$ and $\subtypingd[\Gamma'] S U$, then $\subtypingd S U$. 
\end{theorem}

Given these results, we can proceed to proving the soundness of strong kernel $\dsub$ with respect to full $\dsub$, \Cref{thm:skdsub2dsub}. We prove a stronger result:
\begin{theorem}
  If $\subtypingdsk S U$, $\opesubd{\Gamma}[\Gamma_1]$ and
  $\opesubd{\Gamma}[\Gamma_2]$, then $\subtypingd S U$. 
\end{theorem}
\begin{proof}
  By induction on the strong kernel subtyping derivation.
\end{proof}

Then \Cref{thm:skdsub2dsub} follows from reflexivity of $\opesub$.

Since we will show that strong kernel $\dsub$ is decidable, it cannot also be complete with respect to full
$\dsub$. One example is bad bounds. For example, we are still not able to admit the
following judgment which definitely requires bad bounds.
\begin{mathpar}
  \subtypingd[x : \{A : \top .. \bot\}]\top \bot
\end{mathpar}
This shows that full $\dsub$ is strictly more expressive than strong kernel $\dsub$.

Even if we remove the \infref{dsub:bb} rule from full $\dsub$, the
\infref{dsub:sk:all} rule is strictly weaker than the full \infref{dsub:all} rule. For
example, the following is a derivation in full $\dsub$:
\begin{mathpar}
  \inferrule*[right=\infref{dsub:all}]
  {\inferrule*[right=\infref{dsub:bnd}]
    {\text{straightforward}}
    {\subtypingd[]{\{A : \bot .. \bot\}}{\{A : \bot .. \top\}}} \\
  \inferrule*[right=\infref{dsub:sel2}]
  {\text{straightforward}}
  {\subtypingd[x : \{A : \bot .. \bot\}]{x.A}{\bot}}}
  {\subtypingd[]{\forall(x : \{A : \bot .. \top\})x.A}{\forall(x : \{A : \bot .. \bot\})\bot}}
\end{mathpar}
This judgment is rejected by strong kernel $\dsub$ because the comparison of the returned
types relies on the parameter type to the right of $<:$, which is not possible in
strong kernel $\dsub$. 
Notice that this example uses aliasing information from the right parameter type (i.e. that
$x.A$ is an alias of $\bot$) to reason about the left return type (i.e. that $x.A$ is a subtype
of $\bot$), which is something that strong kernel $\dsub$ cannot do.

On the other hand, strong kernel $\dsub$ \emph{does admit} the motivating aliasing example from the
beginning of this section:
\begin{align*}
  &\text{let }\Gamma = x : \{A : \top .. \top\} \\  
  &\inferrule*[right=\infref{dsub:sk:all}]
    {\inferrule*[right=\infref{dsub:sk:sel1}]
    {\text{reflexivity}}
    {\subtypingdskr[\Gamma]{x.A}{\top}[\Gamma]} \\
  \inferrule*[right=\infref{dsub:sk:top}]
  { }
  {\subtypingdsk[\Gamma ; y : x.A]{\top}{\top}[\Gamma ; y : \top]}}
    {\subtypingdsk[\Gamma]{\forall(y : x.A)\top}{\forall(y : \top)\top}[\Gamma]}
\end{align*}

In general, the \infref{dsub:sk:all} rule admits any subtyping between parameter types
that is admitted by strong kernel. This shows that strong kernel $\dsub$ is strictly
more powerful than kernel $\dsub$.

\subsection{Stare-at subtyping}\label{ssec:stareat}

It remains to show that strong kernel $\dsub$ is decidable. We will present the
decision procedure first. We will prove some of its properties in the next section, and finally
prove that it is a sound and complete decision procedure for strong kernel $\dsub$ in \Cref{ssec:sc}.
The decision procedure is shown in \Cref{fig:stare-at}. We call it
\emph{stare-at subtyping}, inspired by the notation $\stareat S U$. If
we see $\gg$ and $\ll$ as eyes and $<:$ as a nose, then the notation looks like a
face, and the two eyes are staring at the nose.

\begin{figure}
  \begin{mathpar}
    \inferrule*[right=\infnlabel{SA-Top}{dsub:sa:top}]
    { }
    {\stareat T \top}

    \inferrule*[right=\infnlabel{SA-Bot}{dsub:sa:bot}]
    { }
    {\stareat \bot T}

    \inferrule*[right=\infnlabel{SA-VRefl}{dsub:sa:vrefl}]
    { }
    {\stareat{x.A}{x.A}}

    \inferrule*[right=\infnlabel{SA-Bnd}{dsub:sa:bnd}]
    {\stareatr S {S'} \\\\ \stareat U{U'}}
    {\stareat{\{A : S .. U\}}{\{A : S' .. U'\}}}

    \inferrule*[right=\infnlabel{SA-All}{dsub:sa:all}]
    {\stareatr S {S'} \\\\ \stareat[\Gamma_1 ; x : S] U{U'}[\Gamma_2 ; x : S']}
    {\stareat{\forall(x : S) U}{\forall(x : S') U'}}

    \inferrule*[right=\infnlabel{SA-Sel1}{dsub:sa:sel1}]
    {\dcd[\Gamma_2]{x.A} T \\\\ \stareat S T[\Gamma_2']}
    {\stareat S {x.A}}
    
    \inferrule*[right=\infnlabel{SA-Sel2}{dsub:sa:sel2}]
    {\ucd[\Gamma_1]{x.A} T \\\\ \stareat[\Gamma_1'] T U}
    {\stareat{x.A} U}
  \end{mathpar}

  \caption{Definition of stare-at subtyping}\label{fig:stare-at}
\end{figure}

In the same way as for step subtyping, the stare-at subtypinging algorithm searches for
a derivation using the inference rules, backtracking when necessary. We will prove that
this backtracking terminates (\Cref{thm:stare-at-term}).

Stare-at subtyping generalizes step subtyping by operating on two contexts. One can
think of stare-at subtyping as a collaborative game between two players, Alice and
Bob. Alice is responsible for the context and type to the left of $<:$ or $>:$, while
Bob is responsible for the other side. In particular, Alice and Bob are completely
independent and do not need to see the contexts or types held by their collaborator.
Most of the rules are just
straightforward extensions of the corresponding rules of step subtyping with two
contexts, except for three cases: \infref{dsub:sa:all}, \infref{dsub:sa:sel1} and
\infref{dsub:sa:sel2}.

In the \infref{dsub:sa:all} rule, the parameter types are allowed to be different, so
there is an additional premise that compares the parameter types. This
rule can handle not only the aliasing example, but also cases where $S'$ is a
strict subtype of $S$.  When comparing the return types, Alice and Bob work on their
own extended contexts, so subsequently, if Alice and Bob refer to $x$, they
potentially see $x$ at different types.

Similar to step subtyping, stare-at subtyping relies on another operation to handle
path types which generalizes \exposure: \revealing. \upcast and \downcast are
generalized accordingly to reflect the differences between \exposure and \revealing.
Like in step subtyping, the \infref{dsub:rv:top}, \infref{dsub:u:top} and
\infref{dsub:d:bot} rules only apply when no other rules apply and the three
operations are all total.
\begin{figure}
  \revealing
  \begin{mathpar}
    \inferrule*[right=\infnlabel{Rv-Stop}{dsub:rv:stop}]
    {T \text{ is not a path}}
    {\revealingd T T[\Gamma]}

    \inferrule*[right=\infnlabel{Rv-Top}{dsub:rv:top}*]
    { }
    {\revealingd T \top [\bigcdot]}

    \inferrule*[right=\infnlabel{Rv-Bot}{dsub:rv:bot}]
    {\revealingd[\Gamma_1] T \bot}
    {\revealingd[\Gamma_1 ; x : T ; \Gamma_2]{x.A}\bot[\bigcdot]}

    \inferrule*[right=\infnlabel{Rv-Bnd}{dsub:rv:bnd}]
    {\revealingd[\Gamma_1] T {\{A : S .. U\}} \\ \revealingd[\Gamma_1'] U {U'}}
    {\revealingd[\Gamma_1 ; x : T ; \Gamma_2]{x.A}{U'}[\Gamma_1'']}
  \end{mathpar}

  \upcast / \downcast
  \begin{mathpar}
    \inferrule*[right=\infnlabel{U-Top}{dsub:u:top}*]
    { }
    {\ucd{x.A}\top[\bigcdot]}

    \inferrule*[right=\infnlabel{D-bot}{dsub:d:bot}*]
    { }
    {\dcd{x.A}\bot[\bigcdot]}

    \inferrule*[right=\infnlabel{U-Bot}{dsub:u:bot}]
    {\revealingd[\Gamma_1] T \bot}
    {\ucd[\Gamma_1 ; x : T ; \Gamma_2]{x.A}\bot[\bigcdot]}

    \inferrule*[right=\infnlabel{D-Top}{dsub:d:top}]
    {\revealingd[\Gamma_1] T \bot}
    {\dcd[\Gamma_1 ; x : T ; \Gamma_2]{x.A}\top[\bigcdot]}

    \inferrule*[right=\infnlabel{U-Bnd}{dsub:u:bnd}]
    {\revealingd[\Gamma_1] T {\{A : S .. U\}}}
    {\ucd[\Gamma_1 ; x : T ; \Gamma_2]{x.A}{U}[\Gamma_1']}

    \inferrule*[right=\infnlabel{D-Bnd}{dsub:d:bnd}]
    {\revealingd[\Gamma_1] T {\{A : S .. U\}}}
    {\dcd[\Gamma_1 ; x : T ; \Gamma_2]{x.A}{S}[\Gamma_1']}
  \end{mathpar}

  \caption{Definition of \revealing and new definitions of \upcast and \downcast}\label{fig:dsub-revealing}
\end{figure}

%

\revealing is similar to \exposure in that it finds a non-path supertype of the given
type, and its rules mirror those of \exposure. The difference is that in addition to a
type, \revealing also returns a typing context. The typing context is a prefix of the
input typing context long enough to type any free variables that may occur in the type
that \revealing returns.
This returned prefix context participates in further
subtyping decisions and makes it quite easy to prove termination.

The change from \exposure to \revealing, the extra typing context that
\revealing returns, is not related to the two typing contexts in strong
kernel $\dsub$ and in the stare-at subtyping rules. Instead, it is motivated
by the stare-at subtyping termination proof. We conjecture that if stare-at
subtyping were to use \exposure instead of \revealing, it would still compute the
same subtyping relation and it would still terminate, but proving termination
would be significantly more difficult. The use of \revealing is sufficient for stare-at subtyping
to satisfy the properties that we desire of it (which we will prove in the next two sections),
and it makes the termination proof simpler than it would be with \exposure.

The \upcast and \downcast rules have the same structure as those of step subtyping,
except that they return the typing context that they receive from \revealing. In the
cases where they return $\top$ or $\bot$, they return an empty context because these
types have no free variables. Like in step subtyping, the result types of \upcast and
\downcast are used in the \infref{dsub:sa:sel1} and \infref{dsub:sa:sel2} subtyping
rules. These rules use the shortened typing context that is returned from \upcast or
\downcast in their recursive subtyping premises.


\subsection{Properties of Stare-at Subtyping}

We now move on to prove basic properties of the stare-at subtyping algorithm and its operation.
We focus first on basic lemmas that ensure that the \revealing rules satisfy their intended specification.

%
\begin{lemma} (\revealing gives prefixes)
  If $\revealingd S U$, then $\Gamma'$ is a prefix of $\Gamma$.
\end{lemma}


\begin{lemma} (\revealing returns no path)
  If $\revealingd S U$, then $U$ is not a path type.
\end{lemma}

\begin{lemma} (soundness of \revealing)
  If $\revealingd S U$, then $\subtypingd S U$. 
\end{lemma}

\begin{lemma} (well-formedness condition)
  If $\revealingd S U$, $\Gamma$ is well-formed and $fv(S) \subseteq dom(\Gamma)$, then
  $\Gamma'$ is well-formed and $fv(U) \subseteq dom(\Gamma')$. 
\end{lemma}

All these lemmas can be proved by direct induction.

%
%

\upcast and \downcast have properties similar to \revealing. The proofs are much
simpler because the operations are not even recursive.

\begin{lemma}
  The following all hold.
  \begin{enumerate}
  \item If $\udcd {x.A} T$, then $\Gamma'$ is a prefix of $\Gamma$.
  \item If $\ucd {x.A} T$, then $\subtypingd {x.A} T$. 
  \item If $\dcd {x.A} T$, then $\subtypingd T {x.A}$. 
  \item If $\udcd {x.A} T$, $\Gamma$ is well-formed and $x \in dom(\Gamma)$, then
    $\Gamma'$ is well-formed and $fv(T) \subseteq dom(\Gamma')$.
  \end{enumerate}
\end{lemma}

Now we can proceed to prove soundness of stare-at subtyping with respect to full $\dsub$. 
As before, we prove a stronger result.

\begin{theorem} (soundness of stare-at subtyping)
  If $\stareat S U$, $\opesubd{\Gamma}[\Gamma_1]$ and $\opesubd{\Gamma}[\Gamma_2]$,
  then $\subtypingd S U$. 
\end{theorem}
\begin{proof}
  By induction on the derivation of stare-at subtyping.
\end{proof}

A corollary is that if Alice and Bob begin with the same context, then stare-at subtyping is sound
with respect to full $\dsub$.

\begin{theorem}
  If $\stareat[\Gamma]S U[\Gamma]$, then $\subtypingd S U$.
\end{theorem}

Next, we want to examine the termination of the operations. First we want to make sure
that \revealing terminates as an algorithm.

\begin{lemma}
  \revealing terminates as an algorithm.
\end{lemma}
\begin{proof}
    The measure is the length of the input context (the number of variables in its domain).
\end{proof}

Now we want to examine the termination of stare-at subtyping. We first define the
structural measures for types and contexts.

\begin{definition}
  The measure $M$ of types and contexts is defined by the following equations.
  \begin{align*}
    M(\top) &= 1 \\
    M(\bot) &= 1 \\
    M(x.A) &= 2 \\
    M(\forall(x : S) U) &= 1 + M(S) + M(U) \\
    M(\{A : S .. U\}) &= 1 + M(S) + M(U) \\ \\
    M(\Gamma) &= \sum_{x : T \in \Gamma} M(T)
  \end{align*}
\end{definition}

As we can see, the measure simply counts the syntactic size of types and contexts. We
can show that \revealing does not increase the input measure and \upcast and \downcast
strictly decrease it.

\begin{lemma}
  If $\revealingd S U$, then $M(\Gamma) + M(S) \ge M(\Gamma') + M(U)$.

  If $\udcd{x.A} U$, then $M(\Gamma) + M(x.A) > M(\Gamma') + M(U)$.
\end{lemma}

\begin{theorem}\label{thm:stare-at-term}
  Stare-at subtyping terminates as an algorithm. 
\end{theorem}
\begin{proof}
  The measure is the sum of measures of all inputs: for $\stareat S U$, the
  measure is $M(\Gamma_1) + M(S) + M(U) + M(\Gamma_2)$. Since the measure
  just reflects the syntactic sizes, it is easy to see that it decreases in
  all of the cases other
  than \infref{dsub:sa:sel1} and \infref{dsub:sa:sel2}. These two cases are proven
  by the previous lemma. Notice that the proof is this easy because Alice
  and Bob use the returned contexts from \upcast and \downcast in both cases.
\end{proof}

\subsection{Soundness and Completeness of Stare-at Subtyping}\label{ssec:sc}

In the previous section, we showed that stare-at subtyping terminates and is
sound for full $\dsub$. In this section, we strengthen the soundness proof
to strong kernel $\dsub$, and also prove completeness with respect to strong
kernel $\dsub$, to show that the fragment of full $\dsub$ decided by stare-at
subtyping is exactly strong kernel $\dsub$. Our overall approach will mirror
the proofs from \Cref{ssec:sck} of soundness and completeness of step subtyping with respect kernel $\dsub$.

%
First, we connect \revealing with strong kernel $\dsub$. 

\begin{lemma}
  If $\revealingd[\Gamma_1] S T$ and $\subtypingdsk T U$, then $\subtypingdsk S U$.
\end{lemma}

In this lemma, the $\Gamma_1'$ returned from \revealing is not used in the rest of the
statement. The intuition is that strong kernel does not shrink the context as
\revealing does so $\Gamma_1'$ is irrelevant. 

This is all we need to show that stare-at subtyping is sound with respect to strong
kernel $\dsub$.

\begin{theorem} (soundness of stare-at subtyping w.r.t. strong kernel $\dsub$) 

  If $\stareat S U$, then $\subtypingdsk S U$. 
\end{theorem}
\begin{proof}
  The proof is done by induction on the derivation of stare-at subtyping and it is
  very similar to the one of \Cref{thm:kernel:sound}.
\end{proof}

The completeness proof is slightly trickier, because in the \infref{dsub:sa:sel1} and
\infref{dsub:sa:sel2} cases, Alice and Bob work on prefix contexts in the recursive
calls. In contrast, in the \infref{dsub:sk:sel1} and \infref{dsub:sk:sel2} rules of
strong kernel $\dsub$, the subtyping judgements in the premises use the same full
contexts as the conclusions.  Therefore, we need to make sure that working on smaller
contexts will not change the outcome.

\begin{theorem}\label{thm:truncation} (strengthening of stare-at subtyping) If
  $\stareat[\Gamma_1 ; \Gamma'_1 ; \Gamma''_1] S U[\Gamma_2; \Gamma'_2 ; \Gamma''_2]$,
  $fv(S) \subseteq dom(\Gamma_1 ; \Gamma''_1)$ and
  $fv(U) \subseteq dom(\Gamma_2 ; \Gamma''_2)$, then $\stareat[\Gamma_1 ; \Gamma''_1] S
  U[\Gamma_2; \Gamma''_2]$.
\end{theorem}
\begin{proof}
  By induction on the derivation of stare-at subtyping.
\end{proof}

By taking $\Gamma''_1$ and $\Gamma''_2$ to be empty, we know Alice and Bob are safe to
work on the prefix contexts.

%

Now we can prove the completeness of stare-at subtyping.

\begin{theorem} (completeness of stare-at subtyping w.r.t. strong kernel $\dsub$)

  If $\subtypingdsk S U$, then $\stareat S U$. 
\end{theorem}
\begin{proof}
  The proof is similar to the one of \Cref{thm:completeness-kernel}. We also
  need to strengthen the inductive hypothesis to the following: if $\subtypingdsk S U$
  and this derivation contains $n$ steps, then $\stareat S U$ and if $U$ is of the
  form $\{A : T_1 .. T_2\}$, then $\revealingd[\Gamma_1] S {S'}$, and either
  \begin{enumerate}
  \item $S' = \bot$, or
  \item $S' = \{A : T_1' .. T_2' \}$ for some $T_1'$ and $T_2'$, such that
    \begin{enumerate}
    \item $\stareat {T_1} {T_1'}$ and
    \item $\subtypingdsk{T_2'}{T_2}$, and the number of steps in this derivation is less
      than or equal to $n$. 
    \end{enumerate}
  \end{enumerate}
  
  The \infref{dsub:sk:sel1} and \infref{dsub:sk:sel2} cases are trickier. After
  invoking the inductive hypothesis, due to the well-formedness condition of \upcast and
  \downcast, we apply \Cref{thm:truncation} so that the eventual derivation of
  stare-at subtyping works in prefix contexts.
\end{proof}

Therefore, we conclude that strong kernel and stare-at subtyping are the same
language.

Completeness may seem somewhat surprising since stare-at subtyping truncates the
typing contexts in the \infref{dsub:sa:sel1} and \infref{dsub:sa:sel2} cases while
strong kernel subtyping does not.  Technically, the truncation is justified by
\Cref{thm:truncation}. Intuitively, since the prefixes of the typing contexts cover
the free variables of the relevant type, they do include all of the information
necessary to reason about that type. However, it is important to keep in mind that
this is possible only because we have removed the \infref{dsub:bb} rule.  In a
calculus with the \infref{dsub:bb} rule, it is possible that $\subtyping S U$ is false
in some context $\Gamma$ that binds all free variables of $S$ and $U$, but that if we
further extend the context with some $\Gamma'$, that can make
$\subtyping[\Gamma ; \Gamma'] S U$ true due to new subtyping relationships introduced
in $\Gamma'$ by the \infref{dsub:bb} rule.

\section{Discussion and Related Work}

\subsection{Undecidability of Bad Bounds}

In \Cref{ssec:nf}, we showed that the \infref{dsub:trans} rule and the
\infref{dsub:bb} rule are equivalent in terms of expressiveness, and that $\dsub$ and
$\dsub$ without the \infref{dsub:bb} rule are both undecidable. We also showed that
kernel $\dsub$ is decidable.

Kernel $\dsub$ applies two modifications to $\dsub$: it makes
the parameter types in the \infref{dsub:all} rule identical, and it
removes the \infref{dsub:bb} rule. It is then interesting to ask whether
kernel $\dsub$ with the \infref{dsub:bb} rule is undecidable.  We conjecture that it is,
but we do not know how to prove it.
We expect that the proof will not be straightforward. The first problem is to identify
a suitable undecidable problem to reduce from. Most well-known undecidable problems have a clear
correspondence to Turing machines, which have deterministic execution. On the other
hand, (kernel) $\dsub$ can have multiple derivations witnessing the same
conclusion. Therefore, the second step would be to find a deterministic fragment of
$\dsub$ that is still undecidable due to bad bounds. Indeed, discovering a
deterministic fragment was also the first step of \citet{fsub-undec}.  Given the
complexity of $\dsub$, it is hard even to find the fragment that would achieve
these criteria.

This problem is interesting because it investigates the effects that follow from
supporting the \infref{dsub:bb} rule. Currently, in both kernel and strong kernel $\dsub$, the
\infref{dsub:bb} rule is simply removed. This is consistent with the Scala compiler,
which also does not implement this rule.
  However, is it possible to support a fragment of this rule? We
know that in $\dsub$, the \infref{dsub:trans} rule and the \infref{dsub:bb} rule are
equivalent, so recovering a fragment of bad bounds recovers a fragment of transitivity
as well. Moreover, some uses of the rule are not necessarily bad. Consider the following example:
\begin{mathpar}
  \subtypingd[x : \{A : \bot .. \top\}; y : \{A : \bot .. \top\}; z : \{A : x.A
  .. y.A\}]{x.A}{y.A}
\end{mathpar}
In this judgment, before $z$, $x.A$ and $y.A$ show no particular relation, but $z$
claims that $x.A$ is a subtype of $y.A$. This example does not look as bad as other
bad bounds like the one asserting $\top$ is a subtype of $\bot$, because it is
achievable. It would be nice to find a decidable fragment that supports examples
such as this. Doing so will require
a careful analysis of the
decidability of bad bounds.


\subsection{Related Work}

There has been much work related to proving undecidability under certain settings of
subtyping.  \citet{fsub-undec} presented a chain of reductions from two counter
machines (TCM) to $\fsub$ and showed $\fsub$ undecidable. \citet{Kennedy2006OnDO}
investigated a nominal calculus with variance, modelling the situations in Java, C\#
and Scala, and showed that this calculus is undecidable due to three factors:
contraviarant generics, large class hierarchies, and multiple inheritance.
\citet{Wehr:2009:DSB:1696759.1696773} considered two calculi with existential types,
$\mathcal{EX}_{impl}$ and $\mathcal{EX}_{uplo}$, and proved both to be
undecidable. Moreover, in $\mathcal{EX}_{uplo}$, each type variable has either upper
or lower bounds but not both, so this calculus is related to $\dsub$, but
since no variable has both lower and upper bounds, it does not expose the bad bounds
phenomenon. \citet{Grigore:2017:JGT:3009837.3009871} proved Java generics undecidable by
reducing Turing machines to a fragment of Java with contravariance.

So far, work on the $\DOT$ calculi mainly focused on soundness proofs
\cite{wadlerfest-dot,oopsla-dot,simple-sound-proof}.  \citet{abel-algorithmic}
presented step subtyping as a partial algorithm for
$\DOT$ typing. In this paper, we have shown that the fragment of $\dsub$ typed by step subtyping is kernel $\dsub$.
\citet{ASPINALL2001273} showed a calculus with dependent types and subtyping that is
decidable due to the lack of a $\top$ type. \citet{Greenman:2014:GFP:2666356.2594308}
identified the Material-Shape Separation. This separation describes two different usages
of interfaces, and as long as no interface is used in both ways, the type checking problem
is decidable by a simple algorithm.

The undecidability proof in this paper has been mechanized in 
Agda.  There are other fundamental results on formalizing proofs of undecidability.
\citet{pcp-undec} mechanized undecibility proofs of various well-known undecidable
problems, including the post correspondence problem (PCP), string rewriting (SR) and
the modified post correspondence problem. Their proofs are based on Turing machines.
In contrast, \citet{10.1007/978-3-319-66107-0_13} used a call-by-value lambda calculus
as computational model.  \citet{Forster:2019:CUI:3293880.3294096} proved undecibility
of intuitionistic linear logic by reducing from PCP.

\section{Conclusion}

We have studied the decidability of typing and subtyping of the $\dsub$ calculus and
several of its fragments. 
We first presented a counterexample showing that the previously proposed mapping from $\fsub$
to $\dsub$ cannot be used to prove undecidability of $\dsub$.
We then discovered a normal form for $\dsub$ and proved its equivalence
with the original $\dsub$ formulation.
We used the normal form to prove $\dsub$ subtyping and typing undecidable by reductions
from $\fsubm$.
We defined a kernel version of $\dsub$ by removing the bad bounds subtyping rule and restricting
the subtyping rule for dependent function types to equal parameter types, as in kernel $\fsub$.
We proved kernel $\dsub$ decidable, and showed that it is exactly the fragment of $\dsub$
that is handled by the step subtyping algorithm of \citet{abel-algorithmic}. We defined
strong kernel $\dsub$, a decidable fragment of $\dsub$ that is strictly in between 
kernel $\dsub$ and full $\dsub$ in terms of expressiveness, and in particular permits
subtyping comparison between parameter types of dependent function types. This allows
us to handle type aliases gracefully within the subtyping relation. Finally, we proposed stare-at subtyping as an algorithm for deciding
subtyping in strong kernel $\dsub$.
We have mechanized the proofs of our theoretical results in a combination of Coq and Agda.

\begin{acks}                            
\end{acks}


\begin{thebibliography}{30}


\ifx \showCODEN    \undefined \def \showCODEN     #1{\unskip}     \fi
\ifx \showDOI      \undefined \def \showDOI       #1{#1}\fi
\ifx \showISBNx    \undefined \def \showISBNx     #1{\unskip}     \fi
\ifx \showISBNxiii \undefined \def \showISBNxiii  #1{\unskip}     \fi
\ifx \showISSN     \undefined \def \showISSN      #1{\unskip}     \fi
\ifx \showLCCN     \undefined \def \showLCCN      #1{\unskip}     \fi
\ifx \shownote     \undefined \def \shownote      #1{#1}          \fi
\ifx \showarticletitle \undefined \def \showarticletitle #1{#1}   \fi
\ifx \showURL      \undefined \def \showURL       {\relax}        \fi
\providecommand\bibfield[2]{#2}
\providecommand\bibinfo[2]{#2}
\providecommand\natexlab[1]{#1}
\providecommand\showeprint[2][]{arXiv:#2}

\bibitem[\protect\citeauthoryear{Amin, Gr{\"u}tter, Odersky, Rompf, and
  Stucki}{Amin et~al\mbox{.}}{2016}]%
        {wadlerfest-dot}
\bibfield{author}{\bibinfo{person}{Nada Amin}, \bibinfo{person}{Samuel
  Gr{\"u}tter}, \bibinfo{person}{Martin Odersky}, \bibinfo{person}{Tiark
  Rompf}, {and} \bibinfo{person}{Sandro Stucki}.}
  \bibinfo{year}{2016}\natexlab{}.
\newblock \bibinfo{booktitle}{\emph{The Essence of Dependent Object Types}}.
\newblock \bibinfo{publisher}{Springer International Publishing},
  \bibinfo{address}{Cham}, \bibinfo{pages}{249--272}.
\newblock
\showISBNx{978-3-319-30936-1}
\urldef\tempurl%
\url{https://doi.org/10.1007/978-3-319-30936-1_14}
\showURL{%
\tempurl}


\bibitem[\protect\citeauthoryear{Amin, Moors, and Odersky}{Amin
  et~al\mbox{.}}{2012}]%
        {fool-dot}
\bibfield{author}{\bibinfo{person}{Nada Amin}, \bibinfo{person}{Adriaan Moors},
  {and} \bibinfo{person}{Martin Odersky}.} \bibinfo{year}{2012}\natexlab{}.
\newblock \showarticletitle{Dependent object types}. In
  \bibinfo{booktitle}{\emph{19th International Workshop on Foundations of
  Object-Oriented Languages}}.
\newblock


\bibitem[\protect\citeauthoryear{Amin and Rompf}{Amin and Rompf}{2017}]%
        {defint}
\bibfield{author}{\bibinfo{person}{Nada Amin} {and} \bibinfo{person}{Tiark
  Rompf}.} \bibinfo{year}{2017}\natexlab{}.
\newblock \showarticletitle{Type soundness proofs with definitional
  interpreters}. In \bibinfo{booktitle}{\emph{Proceedings of the 44th {ACM}
  {SIGPLAN} Symposium on Principles of Programming Languages, {POPL} 2017,
  Paris, France, January 18-20, 2017}}. \bibinfo{publisher}{{ACM}},
  \bibinfo{pages}{666--679}.
\newblock
\urldef\tempurl%
\url{http://dl.acm.org/citation.cfm?id=3009866}
\showURL{%
\tempurl}


\bibitem[\protect\citeauthoryear{Amin, Rompf, and Odersky}{Amin
  et~al\mbox{.}}{2014}]%
        {path-types}
\bibfield{author}{\bibinfo{person}{Nada Amin}, \bibinfo{person}{Tiark Rompf},
  {and} \bibinfo{person}{Martin Odersky}.} \bibinfo{year}{2014}\natexlab{}.
\newblock \showarticletitle{Foundations of Path-dependent Types}. In
  \bibinfo{booktitle}{\emph{Proceedings of the 2014 ACM International
  Conference on Object Oriented Programming Systems Languages \& Applications}}
  \emph{(\bibinfo{series}{OOPSLA '14})}. \bibinfo{publisher}{ACM},
  \bibinfo{address}{New York, NY, USA}, \bibinfo{pages}{233--249}.
\newblock
\showISBNx{978-1-4503-2585-1}
\urldef\tempurl%
\url{https://doi.org/10.1145/2660193.2660216}
\showDOI{\tempurl}


\bibitem[\protect\citeauthoryear{Aspinall and Compagnoni}{Aspinall and
  Compagnoni}{2001}]%
        {ASPINALL2001273}
\bibfield{author}{\bibinfo{person}{David Aspinall} {and}
  \bibinfo{person}{Adriana Compagnoni}.} \bibinfo{year}{2001}\natexlab{}.
\newblock \showarticletitle{Subtyping dependent types}.
\newblock \bibinfo{journal}{\emph{Theoretical Computer Science}}
  \bibinfo{volume}{266}, \bibinfo{number}{1} (\bibinfo{year}{2001}),
  \bibinfo{pages}{273 -- 309}.
\newblock
\showISSN{0304-3975}
\urldef\tempurl%
\url{https://doi.org/10.1016/S0304-3975(00)00175-4}
\showDOI{\tempurl}


\bibitem[\protect\citeauthoryear{Barendregt}{Barendregt}{1984}]%
        {barendregt1984lambda}
\bibfield{author}{\bibinfo{person}{H.P. Barendregt}.}
  \bibinfo{year}{1984}\natexlab{}.
\newblock \bibinfo{booktitle}{\emph{The lambda calculus: its syntax and
  semantics}}.
\newblock \bibinfo{publisher}{North-Holland}.
\newblock
\showISBNx{9780444867483}
\showLCCN{84005966}
\urldef\tempurl%
\url{https://books.google.ca/books?id=eMtTAAAAYAAJ}
\showURL{%
\tempurl}


\bibitem[\protect\citeauthoryear{Cardelli, Martini, Mitchell, and
  Scedrov}{Cardelli et~al\mbox{.}}{1994}]%
        {CARDELLI19944}
\bibfield{author}{\bibinfo{person}{L. Cardelli}, \bibinfo{person}{S. Martini},
  \bibinfo{person}{J.C. Mitchell}, {and} \bibinfo{person}{A. Scedrov}.}
  \bibinfo{year}{1994}\natexlab{}.
\newblock \showarticletitle{An Extension of System F with Subtyping}.
\newblock \bibinfo{journal}{\emph{Information and Computation}}
  \bibinfo{volume}{109}, \bibinfo{number}{1} (\bibinfo{year}{1994}),
  \bibinfo{pages}{4 -- 56}.
\newblock
\showISSN{0890-5401}
\urldef\tempurl%
\url{https://doi.org/10.1006/inco.1994.1013}
\showDOI{\tempurl}


\bibitem[\protect\citeauthoryear{Cardelli and Wegner}{Cardelli and
  Wegner}{1985}]%
        {Cardelli:1985:UTD:6041.6042}
\bibfield{author}{\bibinfo{person}{Luca Cardelli} {and} \bibinfo{person}{Peter
  Wegner}.} \bibinfo{year}{1985}\natexlab{}.
\newblock \showarticletitle{On Understanding Types, Data Abstraction, and
  Polymorphism}.
\newblock \bibinfo{journal}{\emph{ACM Comput. Surv.}} \bibinfo{volume}{17},
  \bibinfo{number}{4} (\bibinfo{date}{Dec.} \bibinfo{year}{1985}),
  \bibinfo{pages}{471--523}.
\newblock
\showISSN{0360-0300}
\urldef\tempurl%
\url{https://doi.org/10.1145/6041.6042}
\showDOI{\tempurl}


\bibitem[\protect\citeauthoryear{Chlipala}{Chlipala}{2013}]%
        {cpdt}
\bibfield{author}{\bibinfo{person}{Adam Chlipala}.}
  \bibinfo{year}{2013}\natexlab{}.
\newblock \bibinfo{booktitle}{\emph{Certified Programming with Dependent Types:
  A Pragmatic Introduction to the Coq Proof Assistant}}.
\newblock \bibinfo{publisher}{The MIT Press}.
\newblock
\showISBNx{0262026651, 9780262026659}


\bibitem[\protect\citeauthoryear{Curien and Ghelli}{Curien and Ghelli}{1990}]%
        {10.1007/3-540-52590-4_45}
\bibfield{author}{\bibinfo{person}{Pierre-Louis Curien} {and}
  \bibinfo{person}{Giorgio Ghelli}.} \bibinfo{year}{1990}\natexlab{}.
\newblock \showarticletitle{Coherence of subsumption}. In
  \bibinfo{booktitle}{\emph{CAAP '90}},
  \bibfield{editor}{\bibinfo{person}{A.~Arnold}} (Ed.).
  \bibinfo{publisher}{Springer Berlin Heidelberg}, \bibinfo{address}{Berlin,
  Heidelberg}, \bibinfo{pages}{132--146}.
\newblock
\showISBNx{978-3-540-47042-7}


\bibitem[\protect\citeauthoryear{Forster, Heiter, and Smolka}{Forster
  et~al\mbox{.}}{2018}]%
        {pcp-undec}
\bibfield{author}{\bibinfo{person}{Yannick Forster}, \bibinfo{person}{Edith
  Heiter}, {and} \bibinfo{person}{Gert Smolka}.}
  \bibinfo{year}{2018}\natexlab{}.
\newblock \showarticletitle{Verification of PCP-Related Computational
  Reductions in Coq}. In \bibinfo{booktitle}{\emph{Interactive Theorem Proving
  - 9th International Conference, ITP 2018, Oxford, UK, July 9-12, 2018}}
  \emph{(\bibinfo{series}{LNCS 10895})}. \bibinfo{publisher}{Springer},
  \bibinfo{pages}{253--269}.
\newblock
\newblock
\shownote{Preliminary version appeared as arXiv:1711.07023.}


\bibitem[\protect\citeauthoryear{Forster and Larchey-Wendling}{Forster and
  Larchey-Wendling}{2019}]%
        {Forster:2019:CUI:3293880.3294096}
\bibfield{author}{\bibinfo{person}{Yannick Forster} {and}
  \bibinfo{person}{Dominique Larchey-Wendling}.}
  \bibinfo{year}{2019}\natexlab{}.
\newblock \showarticletitle{Certified Undecidability of Intuitionistic Linear
  Logic via Binary Stack Machines and Minsky Machines}. In
  \bibinfo{booktitle}{\emph{Proceedings of the 8th ACM SIGPLAN International
  Conference on Certified Programs and Proofs}} \emph{(\bibinfo{series}{CPP
  2019})}. \bibinfo{publisher}{ACM}, \bibinfo{address}{New York, NY, USA},
  \bibinfo{pages}{104--117}.
\newblock
\showISBNx{978-1-4503-6222-1}
\urldef\tempurl%
\url{https://doi.org/10.1145/3293880.3294096}
\showDOI{\tempurl}


\bibitem[\protect\citeauthoryear{Forster and Smolka}{Forster and
  Smolka}{2017}]%
        {10.1007/978-3-319-66107-0_13}
\bibfield{author}{\bibinfo{person}{Yannick Forster} {and} \bibinfo{person}{Gert
  Smolka}.} \bibinfo{year}{2017}\natexlab{}.
\newblock \showarticletitle{Weak Call-by-Value Lambda Calculus as a Model of
  Computation in Coq}. In \bibinfo{booktitle}{\emph{Interactive Theorem
  Proving}}, \bibfield{editor}{\bibinfo{person}{Mauricio Ayala-Rinc{\'o}n}
  {and} \bibinfo{person}{C{\'e}sar~A. Mu{\~{n}}oz}} (Eds.).
  \bibinfo{publisher}{Springer International Publishing},
  \bibinfo{address}{Cham}, \bibinfo{pages}{189--206}.
\newblock
\showISBNx{978-3-319-66107-0}


\bibitem[\protect\citeauthoryear{Greenman, Muehlboeck, and Tate}{Greenman
  et~al\mbox{.}}{2014}]%
        {Greenman:2014:GFP:2666356.2594308}
\bibfield{author}{\bibinfo{person}{Ben Greenman}, \bibinfo{person}{Fabian
  Muehlboeck}, {and} \bibinfo{person}{Ross Tate}.}
  \bibinfo{year}{2014}\natexlab{}.
\newblock \showarticletitle{Getting F-bounded Polymorphism into Shape}.
\newblock \bibinfo{journal}{\emph{SIGPLAN Not.}} \bibinfo{volume}{49},
  \bibinfo{number}{6} (\bibinfo{date}{June} \bibinfo{year}{2014}),
  \bibinfo{pages}{89--99}.
\newblock
\showISSN{0362-1340}
\urldef\tempurl%
\url{https://doi.org/10.1145/2666356.2594308}
\showDOI{\tempurl}


\bibitem[\protect\citeauthoryear{Grigore}{Grigore}{2017}]%
        {Grigore:2017:JGT:3009837.3009871}
\bibfield{author}{\bibinfo{person}{Radu Grigore}.}
  \bibinfo{year}{2017}\natexlab{}.
\newblock \showarticletitle{Java Generics Are Turing Complete}. In
  \bibinfo{booktitle}{\emph{Proceedings of the 44th ACM SIGPLAN Symposium on
  Principles of Programming Languages}} \emph{(\bibinfo{series}{POPL 2017})}.
  \bibinfo{publisher}{ACM}, \bibinfo{address}{New York, NY, USA},
  \bibinfo{pages}{73--85}.
\newblock
\showISBNx{978-1-4503-4660-3}
\urldef\tempurl%
\url{https://doi.org/10.1145/3009837.3009871}
\showDOI{\tempurl}


\bibitem[\protect\citeauthoryear{Kennedy and Pierce}{Kennedy and
  Pierce}{2006}]%
        {Kennedy2006OnDO}
\bibfield{author}{\bibinfo{person}{Andrew~J. Kennedy} {and}
  \bibinfo{person}{Benjamin~C. Pierce}.} \bibinfo{year}{2006}\natexlab{}.
\newblock \showarticletitle{On Decidability of Nominal Subtyping with
  Variance}.
\newblock


\bibitem[\protect\citeauthoryear{Martin}{Martin}{2010}]%
        {martin2010introduction}
\bibfield{author}{\bibinfo{person}{J. Martin}.}
  \bibinfo{year}{2010}\natexlab{}.
\newblock \bibinfo{booktitle}{\emph{Introduction to Languages and the Theory of
  Computation}}.
\newblock \bibinfo{publisher}{McGraw-Hill Education}.
\newblock
\showISBNx{9780073191461}
\showLCCN{2009040831}
\urldef\tempurl%
\url{https://books.google.ca/books?id=arluQAAACAAJ}
\showURL{%
\tempurl}


\bibitem[\protect\citeauthoryear{Nieto}{Nieto}{2017}]%
        {abel-algorithmic}
\bibfield{author}{\bibinfo{person}{Abel Nieto}.}
  \bibinfo{year}{2017}\natexlab{}.
\newblock \showarticletitle{Towards Algorithmic Typing for DOT (Short Paper)}.
  In \bibinfo{booktitle}{\emph{Proceedings of the 8th ACM SIGPLAN International
  Symposium on Scala}} \emph{(\bibinfo{series}{SCALA 2017})}.
  \bibinfo{publisher}{ACM}, \bibinfo{address}{New York, NY, USA},
  \bibinfo{pages}{2--7}.
\newblock
\showISBNx{978-1-4503-5529-2}
\urldef\tempurl%
\url{https://doi.org/10.1145/3136000.3136003}
\showDOI{\tempurl}


\bibitem[\protect\citeauthoryear{Odersky, Cremet, R\"ockl, and Zenger}{Odersky
  et~al\mbox{.}}{2003}]%
        {nuobj}
\bibfield{author}{\bibinfo{person}{Martin Odersky}, \bibinfo{person}{Vincent
  Cremet}, \bibinfo{person}{Christine R\"ockl}, {and} \bibinfo{person}{Matthias
  Zenger}.} \bibinfo{year}{2003}\natexlab{}.
\newblock \showarticletitle{A Nominal Theory of Objects with Dependent Types}.
  In \bibinfo{booktitle}{\emph{Proc. ECOOP'03}}
  \emph{(\bibinfo{series}{Springer LNCS})}.
\newblock


\bibitem[\protect\citeauthoryear{Pfenning}{Pfenning}{2000}]%
        {PFENNING200084}
\bibfield{author}{\bibinfo{person}{Frank Pfenning}.}
  \bibinfo{year}{2000}\natexlab{}.
\newblock \showarticletitle{Structural Cut Elimination: I. Intuitionistic and
  Classical Logic}.
\newblock \bibinfo{journal}{\emph{Information and Computation}}
  \bibinfo{volume}{157}, \bibinfo{number}{1} (\bibinfo{year}{2000}),
  \bibinfo{pages}{84 -- 141}.
\newblock
\showISSN{0890-5401}
\urldef\tempurl%
\url{https://doi.org/10.1006/inco.1999.2832}
\showDOI{\tempurl}


\bibitem[\protect\citeauthoryear{Pierce}{Pierce}{1991}]%
        {pierce-thesis}
\bibfield{author}{\bibinfo{person}{Benjamin~C. Pierce}.}
  \bibinfo{year}{1991}\natexlab{}.
\newblock \emph{\bibinfo{title}{Programming with intersection types and bounded
  polymorphism}}.
\newblock \bibinfo{thesistype}{Ph.D. Dissertation}. \bibinfo{school}{Carnegie
  Mellon University}.
\newblock


\bibitem[\protect\citeauthoryear{Pierce}{Pierce}{1992}]%
        {fsub-undec}
\bibfield{author}{\bibinfo{person}{Benjamin~C. Pierce}.}
  \bibinfo{year}{1992}\natexlab{}.
\newblock \showarticletitle{Bounded Quantification is Undecidable}. In
  \bibinfo{booktitle}{\emph{Proceedings of the 19th ACM SIGPLAN-SIGACT
  Symposium on Principles of Programming Languages}}
  \emph{(\bibinfo{series}{POPL '92})}. \bibinfo{publisher}{ACM},
  \bibinfo{address}{New York, NY, USA}, \bibinfo{pages}{305--315}.
\newblock
\showISBNx{0-89791-453-8}
\urldef\tempurl%
\url{https://doi.org/10.1145/143165.143228}
\showDOI{\tempurl}


\bibitem[\protect\citeauthoryear{Pierce}{Pierce}{1997}]%
        {fsub-bot}
\bibfield{author}{\bibinfo{person}{Benjamin~C. Pierce}.}
  \bibinfo{year}{1997}\natexlab{}.
\newblock \bibinfo{booktitle}{\emph{Bounded Quantification with Bottom}}.
\newblock \bibinfo{type}{{T}echnical {R}eport} 492.
  \bibinfo{institution}{Computer Science Department, Indiana University}.
\newblock


\bibitem[\protect\citeauthoryear{Pierce}{Pierce}{2002}]%
        {tapl}
\bibfield{author}{\bibinfo{person}{Benjamin~C. Pierce}.}
  \bibinfo{year}{2002}\natexlab{}.
\newblock \bibinfo{booktitle}{\emph{Types and Programming Languages}
  (\bibinfo{edition}{1st} ed.)}.
\newblock \bibinfo{publisher}{The MIT Press}.
\newblock
\showISBNx{0262162091, 9780262162098}


\bibitem[\protect\citeauthoryear{Pierce}{Pierce}{2004}]%
        {atapl}
\bibfield{author}{\bibinfo{person}{Benjamin~C. Pierce}.}
  \bibinfo{year}{2004}\natexlab{}.
\newblock \bibinfo{booktitle}{\emph{Advanced Topics in Types and Programming
  Languages}}.
\newblock \bibinfo{publisher}{The MIT Press}.
\newblock
\showISBNx{0262162288}


\bibitem[\protect\citeauthoryear{Rapoport, Kabir, He, and Lhot\'{a}k}{Rapoport
  et~al\mbox{.}}{2017}]%
        {simple-sound-proof}
\bibfield{author}{\bibinfo{person}{Marianna Rapoport}, \bibinfo{person}{Ifaz
  Kabir}, \bibinfo{person}{Paul He}, {and} \bibinfo{person}{Ond\v{r}ej
  Lhot\'{a}k}.} \bibinfo{year}{2017}\natexlab{}.
\newblock \showarticletitle{A Simple Soundness Proof for Dependent Object
  Types}.
\newblock \bibinfo{journal}{\emph{Proc. ACM Program. Lang.}}
  \bibinfo{volume}{1}, \bibinfo{number}{OOPSLA}, Article
  \bibinfo{articleno}{46} (\bibinfo{date}{Oct.} \bibinfo{year}{2017}),
  \bibinfo{numpages}{27}~pages.
\newblock
\showISSN{2475-1421}
\urldef\tempurl%
\url{https://doi.org/10.1145/3133870}
\showDOI{\tempurl}


\bibitem[\protect\citeauthoryear{Rompf and Amin}{Rompf and Amin}{2016}]%
        {oopsla-dot}
\bibfield{author}{\bibinfo{person}{Tiark Rompf} {and} \bibinfo{person}{Nada
  Amin}.} \bibinfo{year}{2016}\natexlab{}.
\newblock \showarticletitle{Type Soundness for Dependent Object Types (DOT)}.
  In \bibinfo{booktitle}{\emph{Proceedings of the 2016 ACM SIGPLAN
  International Conference on Object-Oriented Programming, Systems, Languages,
  and Applications}} \emph{(\bibinfo{series}{OOPSLA 2016})}.
  \bibinfo{publisher}{ACM}, \bibinfo{address}{New York, NY, USA},
  \bibinfo{pages}{624--641}.
\newblock
\showISBNx{978-1-4503-4444-9}
\urldef\tempurl%
\url{https://doi.org/10.1145/2983990.2984008}
\showDOI{\tempurl}


\bibitem[\protect\citeauthoryear{Team}{Team}{2019}]%
        {agda}
\bibfield{author}{\bibinfo{person}{Agda Team}.}
  \bibinfo{year}{2019}\natexlab{}.
\newblock \bibinfo{title}{Agda 2.5.4.2}.
\newblock
\newblock


\bibitem[\protect\citeauthoryear{Team}{Team}{2018}]%
        {coq}
\bibfield{author}{\bibinfo{person}{The Coq~Development Team}.}
  \bibinfo{year}{2018}\natexlab{}.
\newblock \bibinfo{title}{The Coq Proof Assistant, version 8.8.0}.
\newblock
\newblock
\urldef\tempurl%
\url{https://doi.org/10.5281/zenodo.1219885}
\showDOI{\tempurl}


\bibitem[\protect\citeauthoryear{Wehr and Thiemann}{Wehr and Thiemann}{2009}]%
        {Wehr:2009:DSB:1696759.1696773}
\bibfield{author}{\bibinfo{person}{Stefan Wehr} {and} \bibinfo{person}{Peter
  Thiemann}.} \bibinfo{year}{2009}\natexlab{}.
\newblock \showarticletitle{On the Decidability of Subtyping with Bounded
  Existential Types}. In \bibinfo{booktitle}{\emph{Proceedings of the 7th Asian
  Symposium on Programming Languages and Systems}}
  \emph{(\bibinfo{series}{APLAS '09})}. \bibinfo{publisher}{Springer-Verlag},
  \bibinfo{address}{Berlin, Heidelberg}, \bibinfo{pages}{111--127}.
\newblock
\showISBNx{978-3-642-10671-2}
\urldef\tempurl%
\url{https://doi.org/10.1007/978-3-642-10672-9_10}
\showDOI{\tempurl}


\end{thebibliography}

\nocite{*}

%

\end{document}